\documentclass[aps,prd,floatfix,nofootinbib,12pt]{revtex4-2}
\usepackage{amsfonts}
\usepackage{amsmath}
\usepackage{graphicx}
\usepackage{cancel}
\usepackage{braket}
\usepackage[footnotesize]{caption}
\usepackage{subcaption}
\usepackage[colorlinks,citecolor=blue,urlcolor=blue,linkcolor=black]{hyperref}
\usepackage[font=small,labelfont=bf]{caption}
\begin{document} 
	
\title{Superradiant black hole rocket}
\author{Lucas Acito}
\email{acitolucas@iflp.unlp.edu.ar}
\affiliation{Instituto de F\'\i sica La Plata - CONICET and Departamento de F\'\i sica UNLP,\\ C.C. 67, 1900 La Plata, Argentina}
\author{Nicol\'as E. Grandi}
\email{grandi@fisica.unlp.edu.ar}
\affiliation{Instituto de F\'\i sica La Plata - CONICET and Departamento de F\'\i sica UNLP,\\ C.C. 67, 1900 La Plata, Argentina}
\author{Pablo Pisani}
\email{pisani@fisica.unlp.edu.ar}
\affiliation{Instituto de F\'\i sica La Plata - CONICET and Departamento de F\'\i sica UNLP,\\ C.C. 67, 1900 La Plata, Argentina}

\begin{abstract}
We calculate the total thrust resulting from the interaction between charged scalar modes and a superradiant Reissner-Nordström black hole, when  the modes are deflected by a hemispherical perfect mirror located at a finite distance from the black hole's horizon.
\end{abstract}

\maketitle


\section{Introduction}
 
The superradiance phenomenon in black hole physics refers to the emergence of a particle from a scattering process with more energy than that of the incident state. This implies that energy can be extracted from a black hole. 

However, the above description raises two important questions. First, what happens to the one-way nature of the black hole horizon? The answer is that energy is extracted from the black hole in plane wave states that do not contain information. Second, as energy is taken from the black hole, how do we ensure that its entropy does not decrease? The answer lies in the fact that superradiance only occurs when there is a second charge which allows the horizon area to increase as the energy is removed \cite{Brito}.

Initially, superradiance was described in the context of Kerr black holes, where angular momentum serves as the second charge \cite{Penrose}. However, it can also be observed in electrically charged Reissner-Nordström black holes \cite{Bekenstein}. The advantage here is that these black holes are spherically symmetric, making the analysis of wave modes around them simpler.

The purpose of this note is to investigate whether the superradiant effect can be harnessed to extract momentum from a black hole and generate thrust.

\newpage


\section{The setup}
\label{sec:setup}
\begin{figure}[h]
\centering
\includegraphics[width=.45\textwidth]{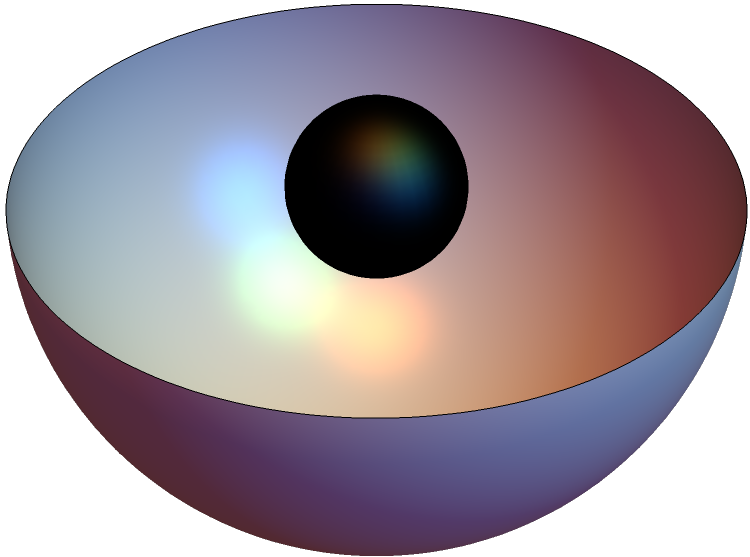}
\caption{A hemispherical perfect mirror around a Reissner-Nordstr\"om black hole.}
\label{fig:rocket}
\end{figure}

The idea we want to test is whether the superradiant phenomenon, which allows us to extract energy from a black hole, can be also used to extract momentum. To keep the calculations as simple as possible, we concentrate in a spherically symmetric charged black hole. In contact with a thermal bath, the black hole superradiates low energy charged particles until equilibrium is reached. To collect the momentum, we use a semispherical mirror centered at the black hole, as shown in Fig. \ref{fig:rocket}. 

To make this proposal concrete, we set a Reissner-Nordstr\"om black hole in asymptotically flat space as a background solution. The bath is made of probe plane wave modes of a charged massive scalar field, featuring a thermal energy distribution at finite chemical potential. The scalar modes are scattered by the black hole and reflected at the mirror, ultimately leading to the total thrust that we aim to compute.

In order to do that, we first solve the classical scattering problem of a scalar plane wave hitting the black hole and mirror system. Then, putting the scalar on a thermal atmosphere corresponds to averaging on the plane wave directions isotropically, and on its energy with a thermal Bose-Einstein distribution.

The total thrust can be then calculated as the flux of the scalar energy momentum tensor on a sphere centered at the black hole and enclosing the mirror.

~ 

\newpage

\section{Scalar modes on a Reissner-Nordstr\"om black hole}\label{sec:background}

Reissner-Nordstr\"om black holes are charged solutions to the Einstein-Maxwell field equations, which have a metric and an electromagnetic field with the form \cite{RN}
\begin{eqnarray}
ds^2&=& -f\, dt^2+\frac{1}{f}\,dr^2+r^2d\Omega_2^2
\qquad\qquad\quad
A=h\,dt
\label{eq:backgound}
\end{eqnarray}
here the lapse $f(r)$ and the electrostatic potential $h(r)$ are functions of $r$ given by
\begin{eqnarray}
f&=&1-\frac{2M%
}{r}+\frac{Q^2}{r^2}
\qquad\qquad\quad
h=-\frac{Q}{r}+\mu
\label{eq:MaxwellandMetricSolution}
\end{eqnarray}
The integration constants $M$ and $Q$ correspond to the ADM mass and charge respectively. Nonextremal solutions $M>Q$ generically have two horizons that we denote $r=r_\pm$. 
  
The equation for a charged Klein-Gordon field with charge ${\sf e}$ and mass ${\sf m}$ in the above background reads
%
%
%
\begin{equation}\label{eq:EOMbackground}
    \frac{1}{r^2}\partial_{r}\left(r^2f \partial_{r}\Phi\right)
    +\left(\frac{1}{r^2}\nabla_{\Omega}^2
    -\frac{\left(\partial_t-i{\sf e}h \right)^2}{f}
    -{\sf m}^2
    \right)\Phi=0
\end{equation}
where $\nabla_\Omega^2$ is the Laplace operator on the two-sphere, with eigenvalues $-\ell(\ell\!+\!1)$ and eigenfunctions given by the spherical harmonics $Y_\ell^m(\theta,\phi) = \sqrt{(2\ell+1)/4\pi}\sqrt{(\ell-m)!/{(\ell+m)!}}\,
$ $e^{i m\phi}P_\ell^m(\cos\theta)$  where $P_\ell^m$ is the associated Legendre polynomial. If we consider a solution with energy $\omega$ we can decompose the scalar field as
\begin{equation}
\Phi(r,\theta,\phi,t)=e^{-i\omega t}\,\sum_{\ell ,m}\alpha^m_{\ell}
\:R_{\ell}(r)\:Y^m_\ell(\theta, \phi)
\label{eq:Decomposition}
\end{equation}
Where $\alpha^m_{\ell}$ are coefficients that depends on $\omega$ which are determined once the boundary conditions are imposed. Plugging back into the equations of motion, we get an equation for the radial dependence in the form
\begin{equation}\label{eq:EOMradial}
    \frac{1}{r^2}\partial_{r}\left(r^2f\partial_{r}R_{\ell}\right)
    +\left(\frac{\left(\omega+{\sf e}\,h\right)^2}{f} -\frac{\ell(\ell+1)}{r^2} -{\sf m}^2
    \right)R_{\ell}=0
\end{equation}
Notice that the equation is independent of the spherical index $m$, which was anticipated in \eqref{eq:Decomposition} when we omitted to include it as a label for $R_{\ell }$.

We solve the above equation imposing causal (ingoing) boundary conditions at the outer horizon $r_+$. At infinity, we impose the conditions of a scattering problem: an incident Coulomb wave with energy $\omega$ and propagating in an arbitrary direction $\check  n$, added to the corresponding scattered outgoing spherical wave. The inner and outer solutions are matched at a finite radius $r_{0}>r_+$, where we place a perfect mirror in the southern hemisphere (Fig \ref{fig:rocket}) at which the scalar field vanishes.

\subsection{Boundary conditions at the horizon}\label{sec:horizon_fields}
Close to the outer black hole horizon $r=r_+$, the lapse function can be expanded as $f(r)=4\pi T_{BH}\,(r\!-\!r_+)$ where $T_{BH}=f'(r_+)/4\pi$ is the Hawking temperature \cite{Chang-Young}. The radial equation is then approximated by
\begin{equation}\label{eq:EOMNearHorizon}
\partial_{\log(r-r_+)}^2R_{\ell } +
\left(\frac{\omega-\omega_s}{4\pi T_{BH}}\right)^2 R_{\ell }  \simeq0
\end{equation}
where we defined the superradiant frequency $\omega_s$ as $\omega_s={\sf e}(\mu_{BH}-\mu)$ with $\mu_{BH}=Q/r_+$
%
%
Equation \eqref{eq:EOMNearHorizon} has two linearly independent solutions representing ingoing and outgoing waves. We call $u_\ell(r)$ the solution to the full radial equation \eqref{eq:EOMradial} that close to the black hole horizon behaves as an ingoing wave  
\begin{equation}
u_{\ell }(r)\simeq e^{{i}\kappa\log(r-r_+)}
\qquad 
\quad 
\qquad {\rm with}\quad \kappa=-\frac{\omega -\omega_s}{4\pi T_{BH}}
\label{eq:SolutionNearHorizon}
\end{equation}
This represents a plane wave with group velocity $v_g=\partial \omega /\partial \kappa =- 4\pi T_{BH}$. Since $v_g<0$ then wave packets ({\em i.e.} information) are falling into the black hole horizon. On the other hand the phase velocity is $v_f=\omega/\kappa=-4\pi T_{BH}\omega/(\omega-\omega_s)$.
Then, if $\omega<\omega_s$ we have $v_f>0$ and plane waves ({\em i.e.} energy) are being radiated from the horizon. This condition characterizes a superradiant mode. 

Notice that the superradiant behavior takes place as long as $\omega_s\neq0$ or in other words $\mu\neq\mu_{BH}$. This can be interpreted as the black hole not being at equilibrium with its environment. Indeed, since $A(r_+)=\mu-\mu_{BH}$, the gauge field does not vanish at the horizon, which would imply a conical singularity in the corresponding Euclidean continuation.

Imposing ingoing boundary conditions at the horizon entails writing the solution in the form \eqref{eq:Decomposition} in the region between the horizon and the mirror, choosing the ingoing form for the radial part $R_{\ell}(r)=u_\ell(r)$,
\begin{align}
  \Phi^-(r,\vartheta,\varphi,t)=e^{-i\omega t}\,\sum_{\ell, m} a_\ell^m\frac{u_\ell(r)}{u_\ell(r_{0})}\,Y_\ell^{m}(\theta,\phi)\:,
  \qquad\qquad r_+<r<r_{0}\:.
\label{eq:SolutionInterior}
\end{align}
Here we 
 rescaled the coefficient as $\alpha_{\ell}^m=a_\ell^m/u_\ell(r_{0})$ in order to simplify the forthcoming calculations.

\subsection{Boundary conditions at infinity}\label{sec:infty_fields}
When the radial coordinate goes to infinity, we can expand the radial equation \eqref{eq:EOMradial} as 
\begin{equation}\label{eq:EOMinfinity}
    \frac{1}{r^2}\partial_r\left(r^2\partial_rR_{ \ell}(r)\right)+
    \left(k^2-\frac{2\eta}{r}-\frac{\ell(\ell+1)+\cdots}{r^2}\right)R_{\ell}(r)=0
\end{equation}
with $k^2={(\omega+{\sf e}\mu)^2-{\sf m}^2}$ and $\eta=Q(\omega+{\sf e}\mu)-2M k^2-M{\sf m}^2$. This corresponds to a Coulomb scattering problem, with solutions behaving as outgoing spherical waves \cite{Gaspar}. 
Calling $(-i)^{\ell+1}v_\ell(r)$ to the solution to the full radial equation \eqref{eq:EOMradial} that matches such behavior at infinity, we have
\begin{equation}\label{eq:SolutionScattededInfinity}
    v_\ell(r)
\simeq {
\frac{1}{ (kr)^{1+i\eta}}\,e^{i kr}}
\end{equation}
%
The scattered part of the scalar field can then be written as in \eqref{eq:Decomposition} with the radial solution replaced by its outgoing form  $R_{\ell}(r)=v_\ell(r)$,
\begin{align}
\Phi^{\sf scat}(r,\theta,\phi,t)=e^{-i\omega t}\,
\sum_{\ell,m}b_\ell^m\frac{v_\ell(r)}{v_\ell(r_0)}
\,Y_\ell^{m}(\theta,\phi)
\,, \qquad\qquad r>r_0\,,
\label{eq:SolutionScattered}
\end{align}
where again we have rescaled $\alpha_{\ell}^m=b_\ell^m/{v_\ell(r_0)}$ for future use.

Regarding the incident wave, we consider a plane wave with momentum $\vec k=k\,\check{n}$, with $\omega_\pm=\pm\sqrt{k^2+{\sf m}^2}-{\sf e}\mu$, and amplitude $A_{\vec k}^\pm$, which in the Coulomb background picks a power law factor. It can be decomposed in outgoing and ingoing spherical waves as
\begin{align}\label{eq:SolutionPlana}
\Phi^{\sf inc}(r,\theta,\phi,t)
  &=
\frac{{\cal N}_\pm}{4\pi} A_{\vec k}\,
e^{i(k r\,\check{n}\cdot\check{r}-\omega_\pm t)
}
(kr)^{-i\eta}\left(1-\check n\cdot\check r\right)^{-i\eta}
  \\
  &\simeq
  {-i }{\cal N}_\pm
  A_{\vec k}^\pm
  \sum_{ \ell m}
	Y_\ell^{m*}(\check n)  \,
	Y_\ell^{m}(\theta,\phi)
\left(
\frac{e^{ikr}}{(2kr)^{1+i\eta}} 
-
(-1)^\ell
\frac{e^{-ikr}}{(2kr)^{1-i\eta}}
\right)  e^{-i\omega_\pm t}
\,.\nonumber
\end{align}
%
Now, calling $i^{\ell+1}
w_\ell(r)$ to the solution that behaves as the parenthesis
at infinity, 
%
%
we can write the incident part of the scalar field as
\begin{align}\label{eq:SolutionIncident}
  \Phi^{\sf inc}(r,\theta,\phi,t)=e^{-i\omega_\pm t}\,\sum_{\ell, m}
c_\ell^m\frac{w_\ell(r)}{w_\ell(r_0)}
\,Y_\ell^{m}(\theta,\phi)\,,
  \qquad\qquad r>r_0\,,
\end{align}
with
%
 $ c_\ell^m =
i^\ell\mathcal{N}_{\pm} 
 \,  A_{\vec k}^\pm\, Y_\ell^{m*}(\check n) \,w_\ell(r_0)$.
Notice that at infinity we have $w_\ell(r)\propto\mathfrak{Re}\{v_\ell(r)\}$ and since both $w_\ell(r),\:v_\ell(r)$ solve the same linear 
equation, then this relation holds for all $r$.

The outer solution is a superposition of the incident and the scattered parts.
In consequence, for $r>r_0$ the  solution reads
\begin{align}
  \Phi^+(r,\theta,\phi,t)&= \Phi^{\sf inc}(r,\theta,\phi,t)+  \Phi^{\sf scat}(r,\theta,\phi,t)\nonumber
\\\label{eq:outer}
&=e^{-i\omega_\pm t}\,
 \sum_{\ell, m}
  \left(b_\ell^m\frac{v_\ell(r)}{v_\ell(r_0)}+c_\ell^m\frac{w_\ell(r)}{w_\ell(r_0)}\right)
	Y_\ell^{m}(\theta,\phi)\,,
  \quad\quad r>r_0\,.
\end{align}

\newpage
\subsection{Matching at the mirror}\label{sec:mirror}
The mirror is a hemispherical shell centered at the black hole that sits at a fixed radius at the southern hemisphere. For a perfect mirror the scalar field vanishes at its surface. We then impose that the scalar field must be zero at $r=r_0$ for $\frac \pi 2\leq\theta\leq\pi$,
\begin{equation}\label{eq:mirror}
\Phi(r_0,\theta,\phi,t)=0\qquad\qquad\frac{\pi}{2} \leq\theta\leq\pi\,.
\end{equation}
On the other hand, the wave function is continuous on the sphere $r=r_0$,
\begin{equation}\label{eq:continuity}
    \Phi^+(r_0,\theta,\phi,t)=\Phi^-(r_0,\theta,\phi,t) \qquad\qquad 0\leq\theta\leq\pi\:,
\end{equation}
whereas the radial derivative is continuous only on the northern hemisphere 
\begin{equation}\label{eq:derivativecontinuity}
    \frac{d}{dr}\Phi^+(r,\theta,\phi,t)\arrowvert_{r=r_0}
    =\frac{d}{dr}\Phi^-(r,\theta,\phi,t)\arrowvert_{r=r_0} \qquad\qquad 0\leq\theta\leq\frac{\pi}{2} \:.
\end{equation}
These three matching conditions at $r=r_0$ determine the coefficients $a_\ell^m$ and $b_\ell^m$ in \eqref{eq:SolutionInterior} and \eqref{eq:SolutionScattered} in terms of $c_\ell^m$ in \eqref{eq:SolutionIncident} or, equivalently, the incident amplitudes $A_{\vec k}^\pm$.
%
%
In particular, continuity of the wave function at $r=r_0$ implies
\begin{align}
  b_\ell^m=a_l^m-c^m_\ell
\end{align}
so we just have to determine the coefficients $a^m_\ell$. Since the solution vanishes at the mirror, we have
\begin{align}\label{eq:bc2}
  0=\sum_{\ell m} a_\ell^m\,Y_\ell^m(\theta,\phi)\qquad \text{for} \quad \frac{\pi}{2}\leq\theta\leq\pi\:.
\end{align}
On the other hand, continuity of the radial derivative at $r=r_0$ implies
\begin{align}\label{eq:bc4}
  0=\sum_{\ell,m}\left(\mathcal{H}_\ell\,a_\ell^m-\mathcal{I}_\ell\,c_\ell^m\right)Y_\ell^m(\theta,\phi)
  \qquad\text{for}\quad0\leq\theta\leq\frac{\pi}{2}\:,
\end{align}
where
\begin{align}
  \mathcal{H}_\ell=\frac{u'_\ell(r_0)}{u_\ell(r_0)}-\frac{v'_\ell(r_0)}{v_\ell(r_0)}\,,
  \qquad
  \mathcal{I}_\ell=\frac{w'_\ell(r_0)}{w_\ell(r_0)}-\frac{v'_\ell(r_0)}{v_\ell(r_0)}
  =-\frac{i}{kr_0^2f(r_0)}\,\frac{1}{w_\ell(r_0)v_\ell(r_0)}\,.
\end{align}
In the second equality 
we use the Wronskian definition and the asymptotic value of the functions. Expressions \eqref{eq:bc2} and \eqref{eq:bc4} hold for any $\phi\in(0,2\pi)$ so they immediately lead to
\begin{align}
  0&=\sum_{\ell=|m|}^\infty
  (-1)^{\ell}\,
  a_\ell^m\,\sqrt{(2\ell+1)}\sqrt{\frac{(\ell-m)!}{(\ell+m)!}}\ P_\ell^{m}(x)\,,\label{eq:bc22}\\[2mm]
  0&=\sum_{\ell=|m|}^\infty
  \left(\mathcal{H}_\ell\,a_\ell^m-\mathcal{I}_\ell\,c_\ell^m\right)
  \sqrt{(2\ell+1)}\sqrt{\frac{(\ell-m)!}{(\ell+m)!}}\ P_\ell^{m}(x)\,,\label{eq:bc44}
\end{align}
for each $m\in\mathbb{Z}$ and $0\leq x=\cos\theta\leq1$. Using that the associated Legendre polynomials $P_\ell^m(x)$ and $P_{\ell'}^m(x)$ are orthogonal in the interval $(0,1)$ whenever $\ell-\ell'$ is even, we obtain
\begin{align}
  0&=a^m_\ell-\sum_{\mbox{\tiny$\ell'\geq|m|$}}
  \left(\gamma^m_{\ell\ell'}-\delta_{\ell\ell'}\right)\,a^m_{\ell'}\,,\label{eq:bc22q1}\\[2mm]
  0&=\mathcal{H}_\ell\,a_{\ell}^m-\mathcal{I}_\ell\,c_{\ell}^m
  +\sum_{\mbox{\tiny$\ell'\geq|m|$}}
  \left(\gamma^m_{\ell\ell'}-\delta_{\ell\ell'}\right)
  \left(\mathcal{H}_{\ell'}\,a_{\ell'}^m-\mathcal{I}_{\ell'}\,c_{\ell'}^m\right)\,,\label{eq:bc22q2}
\end{align}
where we have introduced the coefficients
\begin{align}\label{eq:gamma}
&\gamma^m_{\ell\ell'}=
\sqrt{
\frac{(2\ell+1)(2\ell'+1)(\ell-m)!(\ell'-m)!}{(\ell+m)!(\ell'+m)!} }
\mbox{}\times\int_0^1 dx\,P_{\ell}^{m}(x)P_{\ell'}^{m}(x)\,,
\end{align}
which take the particular values $\gamma^m_{\ell\ell'}=0$ for $\ell-\ell'$ even and $\gamma^m_{\ell\ell}=1$\footnote{Had we considered a mirror with the form of a spherical cap or zone
instead of a full hemisphere, we would have to deal with two different
set of $\gamma^m$ matrices, both with nonvanishing elements for $\ell-\ell'$ even.}. Note that with this values the only nonzero components in the sums \eqref{eq:bc22q1} and \eqref{eq:bc22q2} are with $\ell-\ell'$ odd, but the used notation is more convenient. The matrix $\gamma^m$ of coefficients $\gamma^m_{\ell\ell'}$ is real and symmetric and it satisfies $(\gamma^m)^2=2\gamma^m$ and $(\gamma^m-\mathbf{1})^{-1}=\gamma^m-\mathbf{1}$ (where $\mathbf{1}$ is the identity matrix). The set of matrices $\gamma^m$ is even in the index $m$ satisfying $\gamma^{-m}=\gamma^m$.

Replacing expression \eqref{eq:bc22q1} in \eqref{eq:bc22q2} and regrouping the coefficients, the result reads 
\begin{align}
  \sum_{h= |m|}^\infty
  M^m_{\ell h}\,a^m_h=\sum_{\ell'=|m|}^{\infty}c_{\ell'}^m\mathcal{I}_{\ell'}\gamma_{\ell\ell'}^m\:,
\end{align}
valid for any $\ell\geq |m|$ and where 
\begin{equation}\label{eq:M}
    (M_m)_{\ell h}=\left[-\mathcal{H}(\gamma^m-\mathbf{1})-(\gamma^m-\mathbf{1})\mathcal{H}+
\gamma^m\mathcal{H}\gamma^m\right]_{\ell h}
\end{equation}
with $\mathcal{H}$ a diagonal matrix of elements $\mathcal{H}_\ell$. 
As a consequence, we finally have
\begin{align}
  a^m_\ell&=\sum_{\ell'=|m|}^\infty \left(O^m_{\ell \ell'}+\delta_{\ell \ell'}\right)\,c_{\ell'}^m\,,
\qquad\qquad\quad
  b^m_\ell=\sum_{\ell'=|m|}^\infty 
O^m_{\ell\ell'}
  \,c_{\ell'}^m\,,
\label{eq:b.enfuncionde.c}
\end{align}
where
\begin{align}
    O^m_{\ell\ell'}&=-\frac{i}{k\: r_0^2f(r_0)}\frac{1}{w_{\ell'}(r_0)v_{\ell'}(r_0)}\left[M_m^{-1}\gamma^m\right]_{\ell\ell'}-\delta_{\ell \ell'}
  \label{eq:U}
\end{align}
With this, we get the coefficients $a_\ell$ and $b_\ell$ in terms of the functions $u_\ell(r)$, $v_\ell(r)$ and $w_ \ell(r)$, which have to be obtained numerically by solving the radial equation \eqref{eq:EOMradial} with the appropriate boundary conditions \eqref{eq:SolutionNearHorizon}, \eqref{eq:SolutionScattededInfinity} and \eqref{eq:SolutionPlana} respectively, and the matrix elements $\gamma_{\ell\ell'}^m$ of \eqref{eq:gamma}. 

\subsection{Numerical analysis}\label{sec:numerical_rocket}

In this section we analyze the classical field obtained from the previous sections that fulfill all the boundary conditions. To do so, we need to obtain numerically the radial solutions $u_\ell(r)$, $v_\ell(r)$ and $w_\ell(r)$ of the radial equation \eqref{eq:EOMradial}, and compute $\gamma^m_{\ell\ell'}$. These components characterize the field obtained in the whole space, in terms of the incident wave amplitude $A^\pm_{\vec{k}}$ and direction $\check{n}$ which are free parameters. 

In order to obtain numerically $u_\ell(r)$ that satisfies the near horizon behavior \eqref{eq:SolutionNearHorizon}, first we extend the ingoing wave to a finite radius away from the horizon to avoid the singularity, and then solve numerically up to the mirror. On the other hand, $v_\ell(r)$ and $w_\ell(r)$ are obtained numerically by shooting, integrating from some arbitrary initial condition at the mirror up to a distance where the space-curvature effect is negligible; and there fitting with the asymptotic behavior \eqref{eq:SolutionScattededInfinity} and \eqref{eq:SolutionPlana}. 
 For further details, see the open code \cite{github}. Lastly, the matrix of coefficients $\gamma^m_{\ell\ell'}$ are numerical integrals which can be seen in Appendix \ref{appendix:gamma}. 

\begin{figure}[t]
{\centering
\includegraphics[width=0.36\textwidth]{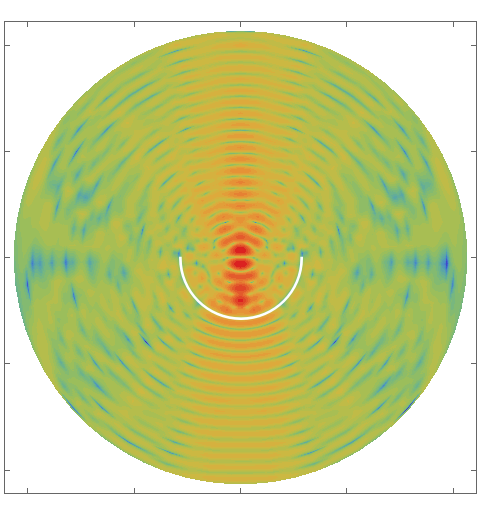}
\hfill
\includegraphics[width=0.486\textwidth]{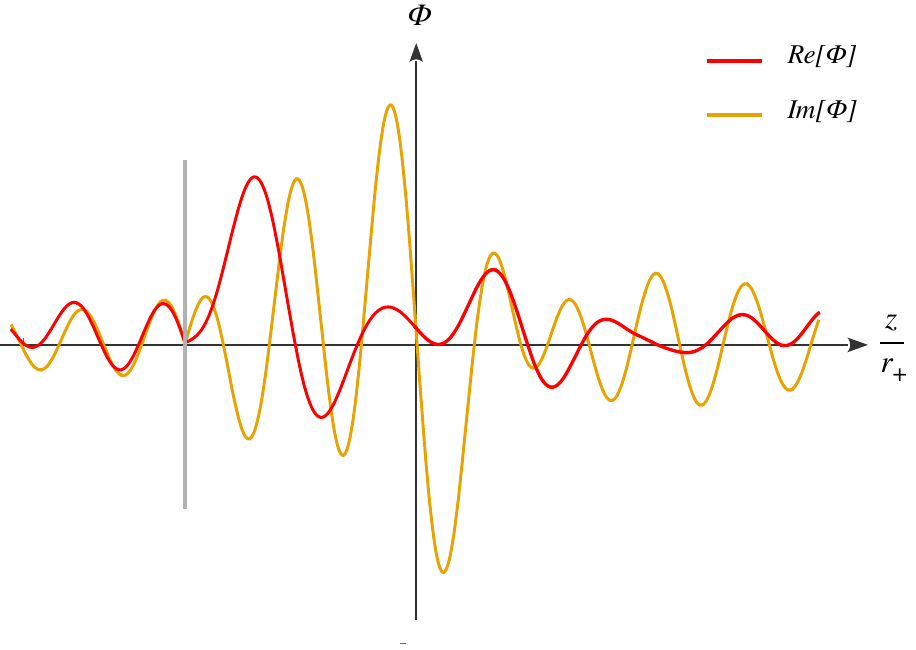}
\\
\includegraphics[width=0.36\textwidth]{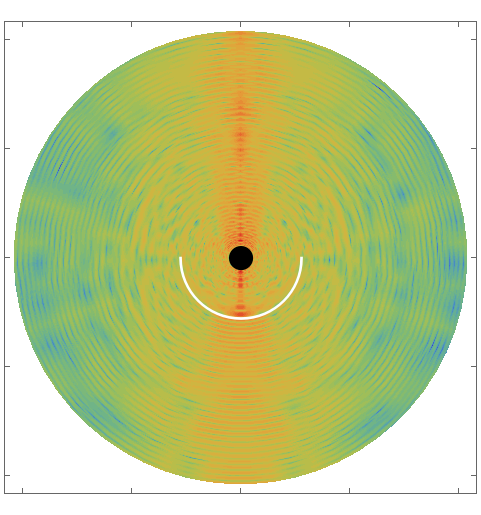}
\hfill 
\includegraphics[width=0.486\textwidth]{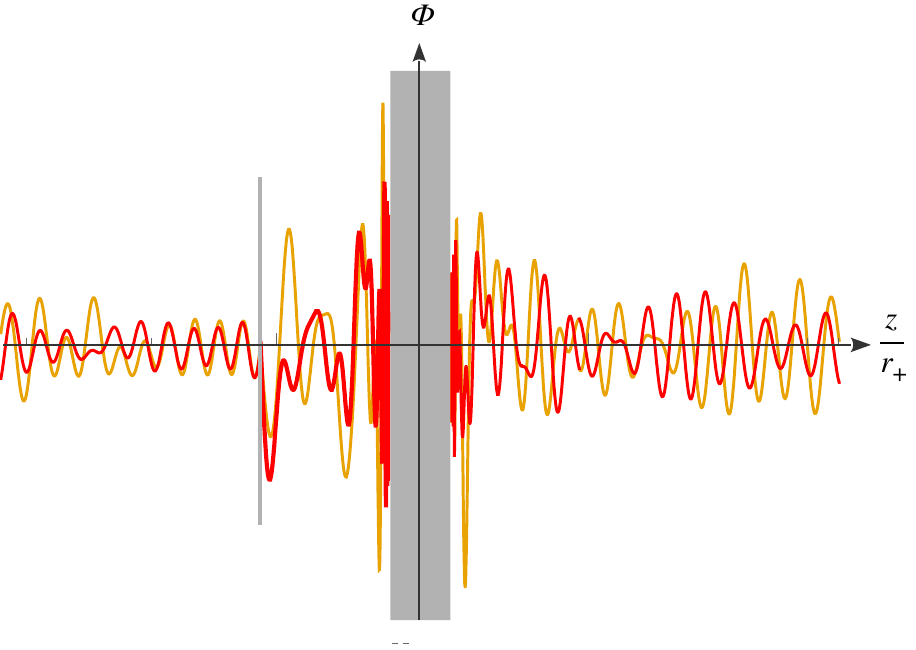}
\\
\includegraphics[width=0.36\textwidth]{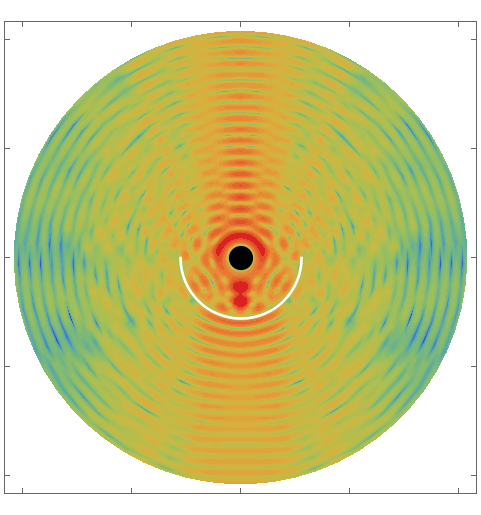}
\hfill
\includegraphics[width=0.486\textwidth]{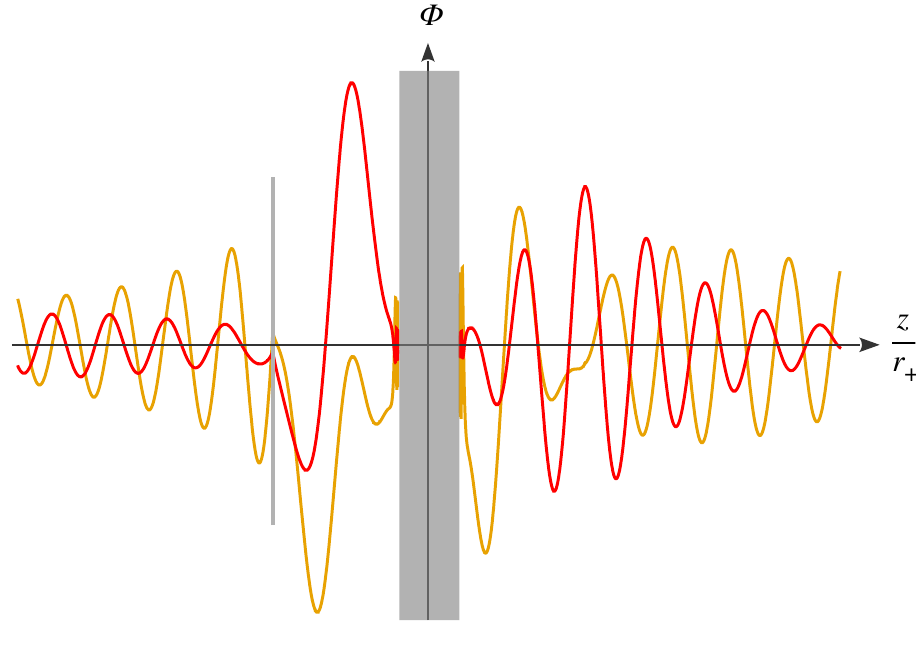}
}
\caption{\small Classical field for a superposition of incident waves with $0\leq\theta\leq\pi$  
with ${\sf m}=0.1$ and ${\sf e}=0.5$. 
\underline{Left:} amplitude $|\Phi|^2$ on a transverse plane containing the $z$ axis. 
\underline{Right:} field $\Phi(0,0,z)$ as a function of $z$. First line: on flat space $M=Q=0$ with $k=0.3$. Second and third lines: on a black hole with $M=10$, $Q=0.99 M$, with $k=0.6$ (nonsuperradiant mode) and $k=0.3$ (superradiant mode) respectively.
\normalsize
\label{fig:field_noBH}}
\end{figure}

To gain some intuition on the resulting field, we first consider the flat space case (see Fig. \ref{fig:field_noBH}, first line) with an isotropic set of incoming waves with equal fixed amplitude $A^\pm_{\vec{k}}$. In this case, the radial solutions are the spherical Bessel functions $(-i)^{\ell+1}u_\ell(r)=(-i)^{\ell+1}w_\ell(r)=j_\ell(kr)$ and the first kind spherical Hankel functions $(-i)^{\ell+1}v_\ell(r)=h_\ell^{(1)}(kr)$.
On the left plot we see the charge density $|\Phi^2|$ on a plane that contains the $z$ axis. We can see the spherical aberration pattern in the form of a high intensity vertical region at the center of the plot, as expected for a hemispherical mirror. The field $\Phi$ as a function of the variable $z$ is depicted on the right, where we  can see how the matching works 
at the mirror radius. 
We can expect that, as particles are coming from every direction, the mirror would move downward, since there are more particles impinging from the concave side of the mirror. As a consequence, 
we have a `background force' acting on the mirror. 

Next we consider the charged black hole background in the second and third lines of figure \ref{fig:field_noBH}. There, we show two different energies, the plots on the second line correspond to the nonsuperradiant regime $\omega>\omega_{s}$, while those on the third have superradiant energies $\omega<\omega_s$. Again, we can see how the matching of the field works in the $z$ direction and also the behavior near the black hole. 
On the left we can see clearly how the black hole rocket works, as the energy going out from the black hole due to superradiance pushes on the mirror. 

\section{Calculation of the thrust}

\subsection{Quantization}
\label{sec:Quantization}
In the previous sections we have written the full solution for a classical scalar field corresponding to an incident plane wave that scatters at the hemispherical mirror and satisfies ingoing conditions at the black hole horizon. 
Next, we will submerge the setting in a bath of charged scalar particles \cite{Greiner} at temperature $T$ and chemical potential $\mu$. 
In order to satisfy the canonical relation $[\Phi(t,\vec{x}),\dot{\Phi}^{\dagger}(t,\vec{y})]= i \delta^{(3)}(\vec{x}-\vec{y})$, we chose the normalization  
%
%
\begin{equation}\label{eq:quantum_field}
    \mathcal{N}_{\pm}= \frac{|\omega_{\pm}+{\sf e}\mu|}{2\pi^2\sqrt{|\omega_+\omega_-|}}\frac{1}{\sqrt{\pm 2\:\omega_\pm}}
\end{equation}
In this way, the operators $\hat A_{\vec k}^\dagger$ and $\hat B_{\vec k}^\dagger\:$, which are the quantum upgrade of the amplitudes $A_{\vec k}^+$ and $A_{\vec k}^-$ of the classical field \eqref{eq:SolutionPlana}, create particles and antiparticles with momentum $\vec k$ and satisfy the usual algebra $[\hat A_{\vec k},\hat A_{\vec k'}^\dagger]=[\hat B_{\vec k},\hat B_{\vec k'}^\dagger]=(2\pi)^3\,\delta(\vec k-\vec k')$.

Now we assume an isotropic thermal distribution of plane waves with temperature $T$ and chemical potential $\mu$. This implies that the expectation values of the quadratic operators $\hat A_{\vec k}^\dagger \hat A_{\vec k'}$ and $\hat B_{\vec k}^\dagger \hat B_{\vec k'}$ are given by the thermal Bose-Einstein distribution, in the form \cite{Kapusta}
\begin{align}
&\label{eq:adaggera}\braket{\hat{A}_{\vec k}^{\dagger}\hat{A}_{\vec k}}=\frac{1}{e^{\frac{\omega_+}{T}}-1}(2\pi)^3\delta^{3}(\vec{k}-\vec{k}')\,,
\qquad\qquad 
\braket{\hat{B}^{\dagger}_{\vec k}\hat{B}_{\vec k}}=\frac{1}{e^{-\frac{\omega_-}{T}}-1}(2\pi)^3\delta^{3}(\vec{k}-\vec{k}')
\,.
\end{align}
Any other expectation value quadratic in those operators is zero.

\subsection{Total thrust}\label{sec:thrust_calculation}

To calculate the thrust induced on the system by the field modes, we use the energy momentum tensor of the scalar field. The flux of any of its spatial components through a spherical surface enclosing the system at infinity gives the total force in the corresponding direction. 
%
%
\begin{equation}
\label{eq:force}
F_j=\int_{r\to\infty} \!d\Omega_2\ n^i_r T_{ij}=
\int_{r\to\infty} \!d\Omega_2\,
\left.\left(
\partial_jr \,T_{rr}
+\partial_j\theta\, T_{r\theta}
+\partial_j\varphi\, T_{r\varphi}
\right)\right.\,,
\end{equation}
where the indices $i,j$ refer to the asymptotic Cartesian coordinates $x^i$, and the unit vector $n^i_r$ points in the radial direction. 

~

In the first term in \eqref{eq:force} we have $\partial_i\,r=x^i/r=(\cos\varphi\sin\theta,$ $\sin\varphi\sin\theta,$ $\cos\theta)$.
Due to the symmetry of the setting, the energy momentum tensor at spatial infinity does not depend on $\varphi$ so the only nonvanishing component of the thrust from the first term in the integrand points in the $z$ direction. Regarding the second term, we have $r\partial_i\theta=(\cos\varphi\cos\theta,\sin\varphi\cos\theta,-\sin\theta)$, so again only the $z$ component remains. For similar reasons, there is no contribution from the third term. We are left with
\begin{equation}\label{eq:thrust_def}
F_j=\delta_{jz}\,2\pi
\int_0^{\pi} \!d\theta \sin\theta
\left. \left(
\cos\theta
\,r^2T_{rr}
-
\sin\theta
\, rT_{r\theta}
\right)\right|_{r\to\infty}
\end{equation}
We use the energy momentum tensor obtained as the mean value of the corresponding quantum expression (see Appendix \ref{appendix:Tmunu}).
%
%
The needed components $T_{rr}$ and $T_{r\vartheta}$ at $r\to\infty$ are
\begin{align}
    T_{rr}&=\frac{1}{2} \left(\partial'_{t}\partial_{t}+i{\sf e}\mu(\partial'_{t}-\partial_{t}) 
    +\partial_{r}\partial'_{r} -{\sf m}^2+{\sf e}^2\mu^2\right) \braket{\,:\!\hat{\Phi}^{\dagger}(\mathbf{x})\hat{\Phi}
	(\mathbf{x'})\!:\,}\arrowvert_{\mathbf{x'}=\mathbf{x}}  
	\\
    T_{r\theta}&=\frac{1}{2} \left(\partial_{r}\partial'_{\theta} +\partial'_{r}\partial_{\theta}\right) 
    \braket{\,:\!\hat{\Phi}^{\dagger}(\mathbf{x})\hat{\Phi}(\mathbf{x'})\!:\,}\arrowvert_{\mathbf{x'}=\mathbf{x}} 
\end{align}
with the colons $:\cdot:$ denoting normal ordering. Here $rT_{r\theta}$ vanishes at infinity and for $r^2T_{rr}$ we have (see Appendix \ref{appendix:Tmunu})
\small
\begin{multline}\label{eq:T_rr}
    r^2T_{rr}= \int_0^{\infty}\! dk \, \frac{|\omega_++{\sf e}\mu|^2k^2}{\pi\,|\omega_+\omega_-|}\left(\frac{1}{\omega_+\left(e^{\frac{\omega_+}{T}}-1\right)} -\frac{1} {\omega_-\left(e^{-\frac{\omega_-}{T}}-1\right)} \right)
    \sum_{\ell,\ell',m}  \mathcal{U}_{\ell\ell'}Y_\ell^{m*}(\theta,\phi)Y_{\ell'}^{m}(\theta,\phi)
\end{multline}
\normalsize
where we defined
\begin{equation}\label{eq:N_ll'}
    \mathcal{U}_{\ell\ell'}^m=\sum_{\ell''=|m|} \: e^{i\frac{\pi}{2} (\ell-\ell')} 
	\frac{|w_{\ell''}(r_0)|^2}{v_{\ell}^*(r_0)v_{\ell'}(r_0)} O_{\ell\ell''}^{m*} O^m_{\ell'\ell''}
	+\mathfrak{Re}\left[e^{i\frac{\pi}{2} (\ell-\ell')}\frac{w_{\ell}(r_0)}{v_{\ell'}(r_0)}O^m_{\ell'\ell}\right]
\end{equation}
Inserting into \eqref{eq:force} and computing the angular integral, we get 
%
\small
\begin{equation}\label{eq:thrust_result}
    \begin{split}
        F_{\sf z}&= \frac{1}{\pi r_0^2f(r_0)} \int_0^{\infty} dk \: 
\frac{|\omega_++{\sf e}\mu|^2k^2}{\pi\,|\omega_+\omega_-|}\left(\frac{1}{\omega_+\left(e^{\frac{\omega_+}{T}}-1\right)} -\frac{1} {\omega_-\left(e^{-\frac{\omega_-}{T}}-1\right)} \right)        
        \times
        \\
        &\times\sum_{m}\sum_{\ell=|m|}\sqrt{\frac{(\ell-m)(\ell+m)}{(2\ell+1)(2\ell-1)}}\,\mathfrak{Re}\!\left[\frac{1}{v_{\ell-1}^*(r_0)v_{\ell}(r_0)}\left( \left[M_m^{-1}\gamma^m\right]_{\ell-1,\ell}^*-\left[M_m^{-1}\gamma^m\right]_{\ell,\ell-1} \right.\right.
        \\&\hspace{6cm}
        \left.\left.+2 i\!\sum_{\ell''=|m|} \frac{\left[M_m^{-1}\gamma^m\right]_{\ell-1,\ell''}^{*}\left[M_m^{-1}\gamma^m\right]_{\ell,\ell''}}{k\:r_0^2f(r_0)\:|v_{\ell''}(r_0)|^2}\right) \right]
    \end{split}
\end{equation}
\normalsize
where $m\in\mathbb{Z}$, $\ell\in\mathbb{N}_0$ and we have changed the overall sign in order to obtain the force acting {on} the system.  

\subsection{Numerical results}\label{sec:numerical_force}
\begin{figure}[t]
    \centering
\includegraphics[width=0.49\linewidth]{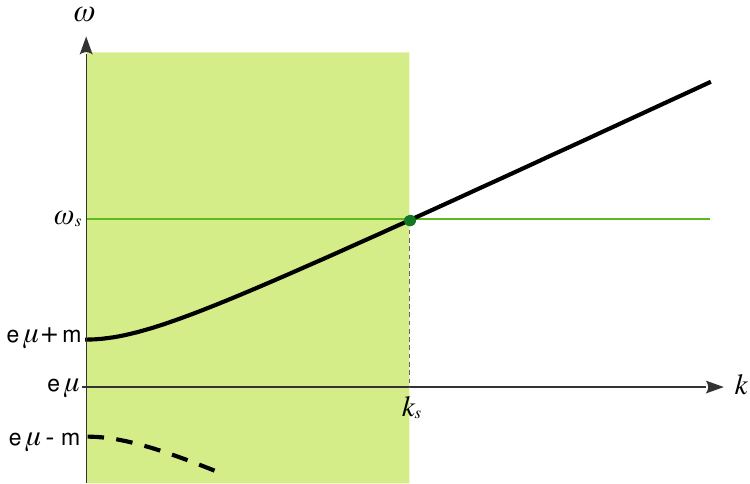}
\hfill
\includegraphics[width=0.49\linewidth]{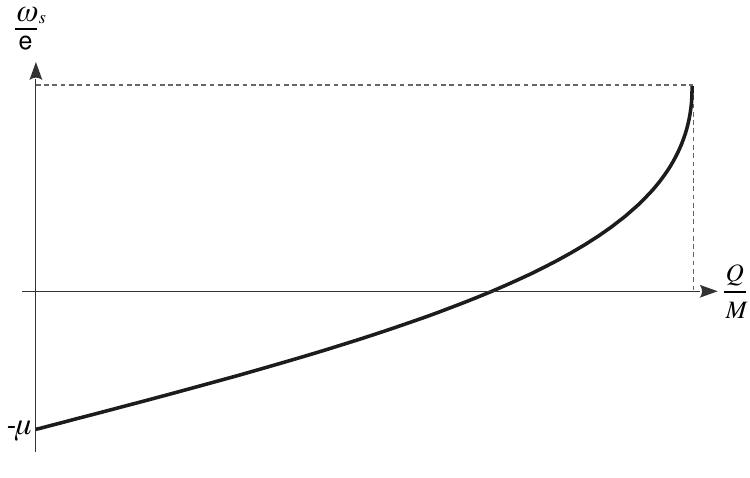}
\caption{Superradiant phenomenon. \underline{Left:} plane wave frequency $\omega_\pm$ and superradiant frequency. 
The green region is the superradiant region and the green dot mark the intersection at $k_s$. \underline{Right:} behavior of the superradiance frequency $\omega_s$ with the ratio $Q/M$ and the charge ${\sf e}$ }
\label{fig:superradiance3}
\end{figure}
 
The numerical analysis of the force \eqref{eq:thrust_result} starts by noticing that it depends on the same classical pieces as in Sec. \ref{sec:numerical_rocket},  {\em i.e.} the inverse operator $M_m^{-1}$, the matrices $\gamma^m$ and the radial solutions evaluated at the mirror radius. 

To understand how superradiance works, we plot in Fig. \ref{fig:superradiance3} (left) the dispersion relation $\omega_\pm=\pm\sqrt{k^2+{\sf m}^2}-{\sf e}\mu$ as compared to the superradiant frequency $\omega_s$. The intersection at $k=k_s$ divides the $k$ axis into two regions: For $k\geq k_s$ we have no superradiance and we expect to have a positive thrust force (pushing the mirror upward). 
For $k\leq k_s$ on the other hand, the system superradiates and thus we expect a negative thrust force (pushing the mirror downward). Moreover, in Fig.~\ref{fig:superradiance3} (right) we plot the superradiant frequency $\omega_s$ as a function of the black hole charge, showing that it is maximal for the extremal black hole, while it goes to $-\mu$ in the Schwarzschild limit. 

\begin{figure}[t]
\centering 
\includegraphics[width=.47 \textwidth]{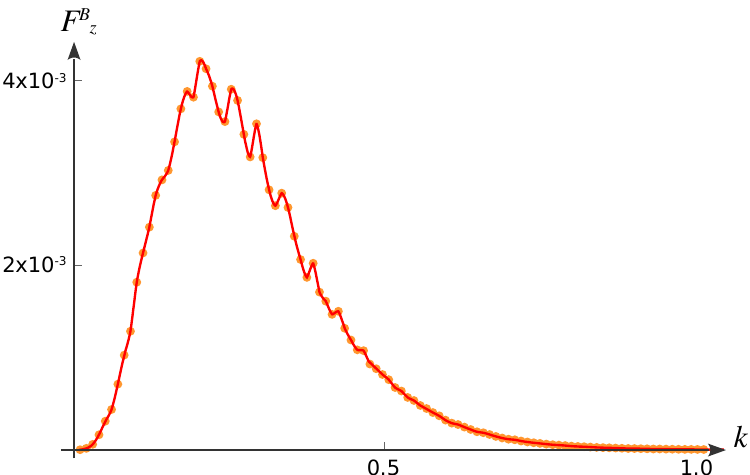}
\caption{Thrust force without the black hole $M=0$, $Q=0$. This plot corresponds to ${\sf m}=0.1$, ${\sf e}=3 {\sf m}$, $\mu=-0.01$, $T=0.07$. The points represent the values of the integrand in \eqref{eq:thrust_result} as a function of $k$, and the resulting thrust force is given the area under the interpolating curve.   
}
\label{fig:background_force}
\end{figure}

We start by calculating the thrust without any black hole, corresponding to an isotropic superposition of modes with momentum $k$. As anticipated in Sec. \ref{sec:numerical_rocket}, we get a nonvanishing background force, see Fig.~\ref{fig:background_force}. This happens because the mirror at rest is not in thermodynamic equilibrium with the thermal bath. As it is accelerated through the bath, the force eventually equilibrates with the drag on the opposite side, resulting in an equilibrium state in which the mirror moves with constant velocity. The background force depends on the   parameters $T$ and $\mu$, being not sensitive to the particle mass ${\sf m}$ or charge ${\sf e}$.    

\begin{figure}[t]
\includegraphics[width=0.47\linewidth]{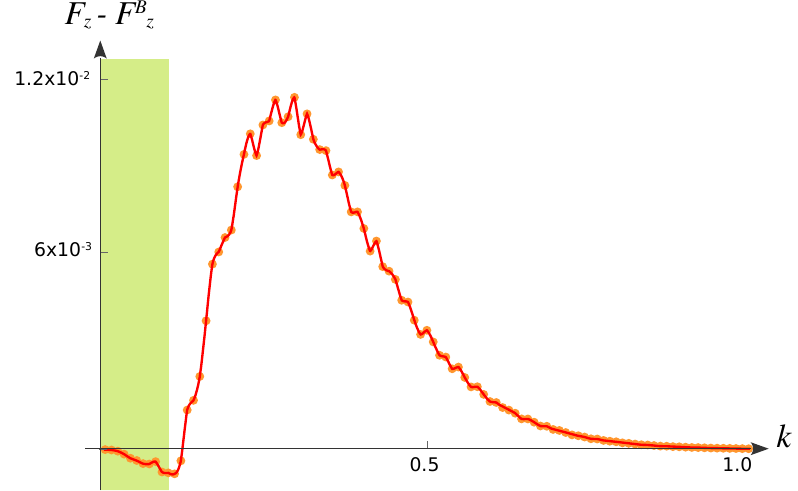}
\hfill
\includegraphics[width=0.47\linewidth]{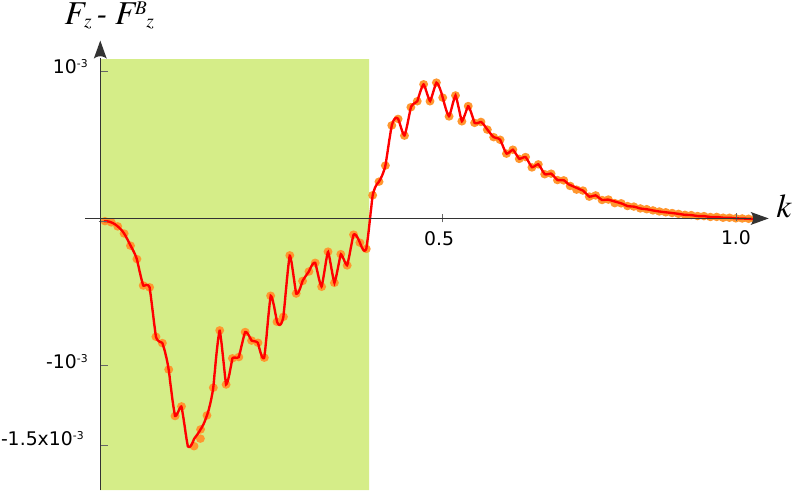}
\caption{Black hole thrust force obtained by removing the background. The green regions are superradiant. Left: we take $Q=0.8M$ and ${\sf e}=3 {\sf m}$ for a smaller superradiant region. Right: we take $Q=0.99M$ and ${\sf e}=5 {\sf m}$ for a bigger superradiant region. These plots correspond to $r_0=5 r_+$, ${\sf m}=0.1$, $\mu=-0.01$, $T=0.07$. The points represent the values of the integrand in \eqref{eq:thrust_result} as a function of $k$, and the resulting thrust force is given the integral under the interpolating curve.  }
    \label{fig:forces}
\end{figure}

We move now to the case in which a charged black hole sits at the centre of the geometry. In Fig.~\ref{fig:forces} we show the thrust force for two different black holes, from which the background contribution has been substracted. We consider one case which is not favorable to superradiance (left) and another one with a large superradiance region (right). As it could have been expected, there is a change of sign in the resulting force  as we move into the superradiance region. These plots correspond to an isotropic superposition of modes with fixed $k$, the total thrust corresponding to the integral over $k$. We see that in the less superradiant case the total area bellow the curve is positive, while in the more superradiant case it is negative.

These results imply that it is possible to use 
the superradiant process to effectively build a ``superradiant black hole rocket''. As we discuss in what follows, the rocket  parameters can be optimized to maximize the resulting thrust. 

\begin{figure}[t]
\centering
\includegraphics[width=0.49\linewidth]{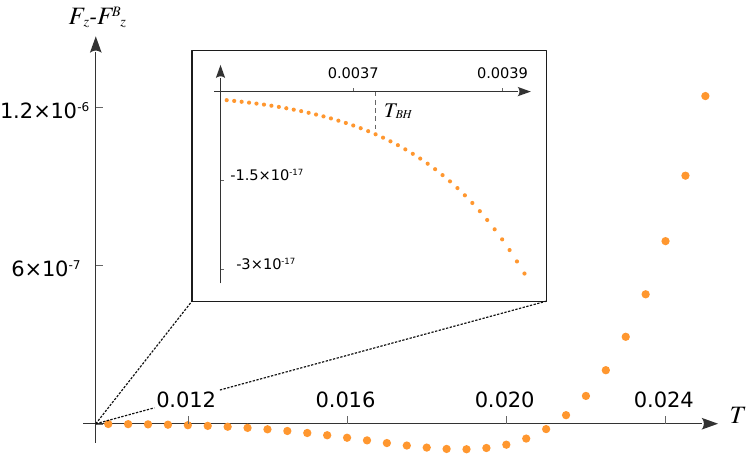}
\hfill
\includegraphics[width=0.49\linewidth]{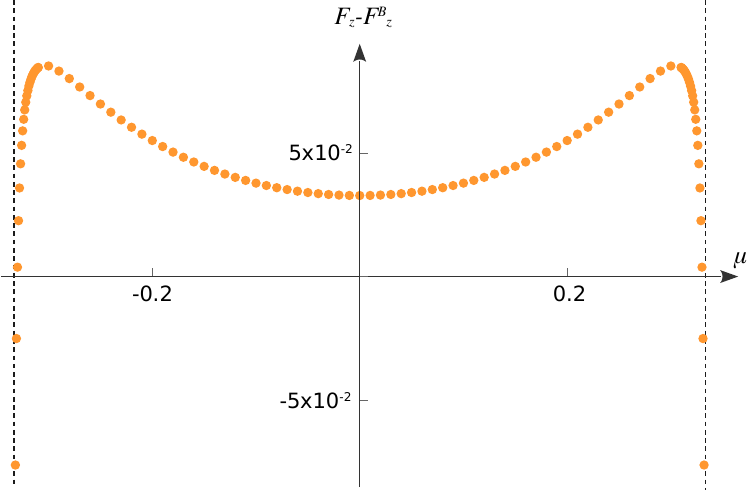}
\caption{Thrust force as a function of the thermal bath parameters, with the same background parameters as in Fig. \ref{fig:forces} (left).}
\label{fig:varying_T_mu}
\end{figure}

The parameters of the thermal bath $T$ and $\mu$ only appear in the force \eqref{eq:thrust_result} through the prefactor containing the Bose-Einstein distribution. This dependence is easy to analyze with the help of the plots in  Fig.~\ref{fig:forces}. Indeed, since the modes with small $k$ have a larger weight for lower the temperatures, we can always get a negative (superradiant) total force. This can be seen in the plot in Fig.~\ref{fig:varying_T_mu} (left). 
On the other hand, the modes with smaller $k$ get a larger weight as the chemical potential approaches its limiting values $\pm|{\sf m}|/{\sf e}$. This results in a change of sign of the total force as the superradiant region become dominant, see the plot in  Fig.~\ref{fig:varying_T_mu} (right). 

\begin{figure}[b]
 \centering
    \includegraphics[width=.6\linewidth]{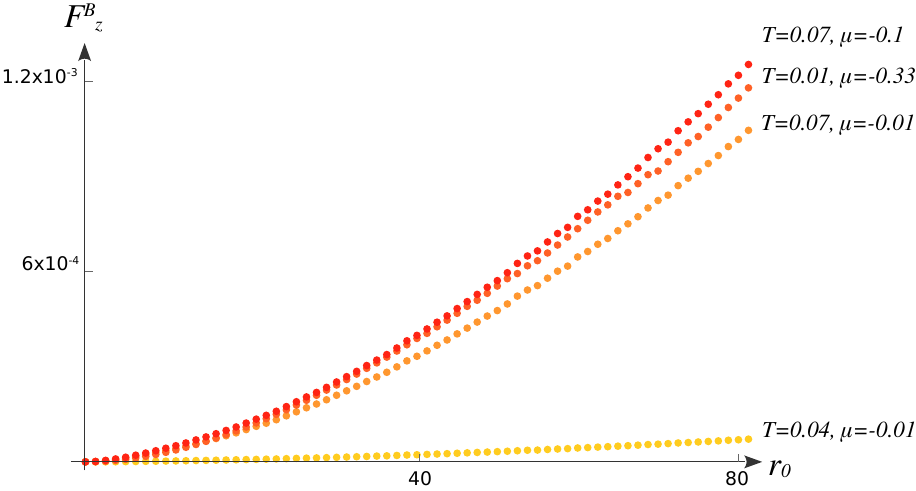}
    \caption{Background force for different thermal baths and mirror radii, from $r_0=0.1$ to $r_0=80$. }
    \label{fig:background_rm_variation}
\end{figure}

\begin{figure}[t]
    \centering
    \begin{subfigure}{.49\textwidth}
        \centering
        \includegraphics[width=\linewidth]{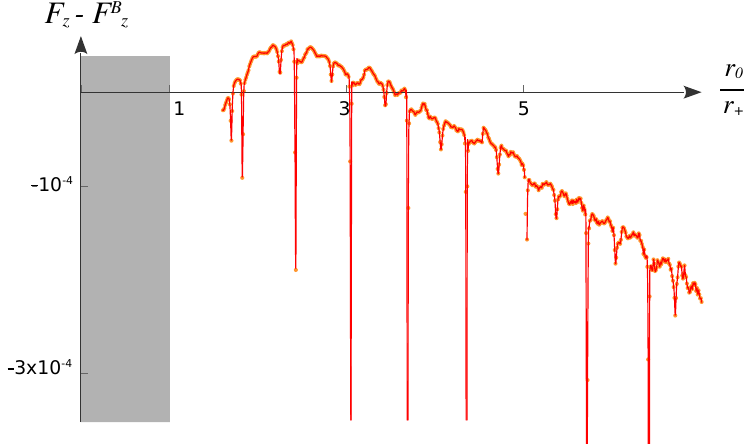}
        \subcaption{$T=0.07, \mu=-0.01$}
        \label{fig:BHforce_rm_1}
    \end{subfigure}%
    \hfill
    \begin{subfigure}{.49	\textwidth}
        \centering
        \includegraphics[width=\linewidth]{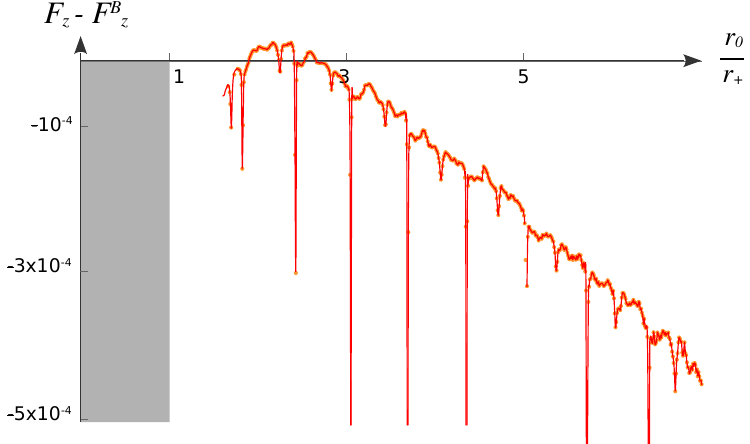}
        \subcaption{$T=0.07, \mu=-0.1$}
        \label{fig:BHforce_rm_2}
    \end{subfigure}
    \begin{subfigure}{.49\textwidth}
        \centering
        \includegraphics[width=\linewidth]{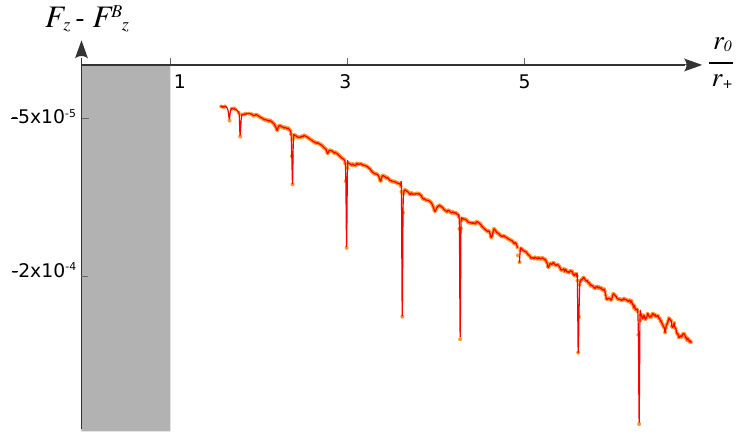}
        \subcaption*{$T=0.05, \mu=-0.1$}
        \label{fig:BHforce_rm_3}
    \end{subfigure}%
    \hfill
    \begin{subfigure}{.49\textwidth}
        \centering
        \includegraphics[width=\linewidth]{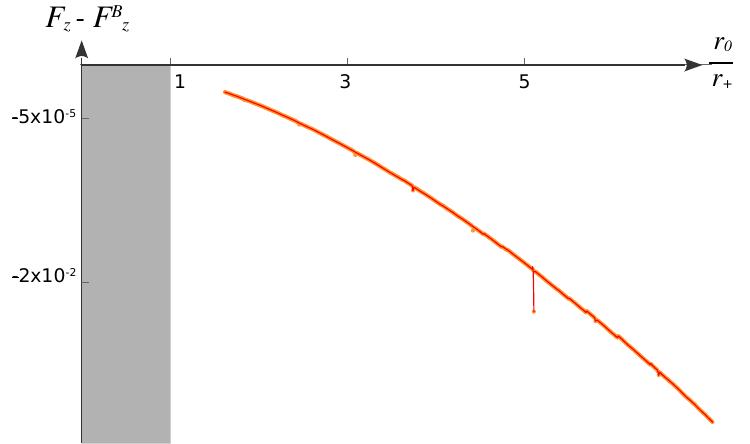}
        \subcaption*{$T=0.05, \mu=-0.2$}
        \label{fig:BHforce_rm_4}
    \end{subfigure}
    \caption{Total thrust force obtained as a function of the mirror radius $r_0$ normalized with the horizon $r_+$ at different $T$ and $\mu$. Here we fix $Q=0.99M$ and ${\sf e}=5{\sf m}$. }
    \label{fig:varying_rm}
\end{figure}

Lastly, we studied the variation of the thrust as a function of the mirror radius $r_0$. We first plot in Fig. \ref{fig:background_rm_variation} the background force without the black hole, as a function of the mirror radius and for different values of the thermal parameters. As expected, it vanishes at $r_0=0$ and grows monotonically with $r_0$. Then, in Fig. \ref{fig:varying_rm} we obtain the thrust force in the presence of the black hole, varying the mirror radius $r_0$ for different values of the temperature $T$ and chemical potential $\mu$.  We see that the sign of the force can change with the radius and, more interestingly,  a set of peaks appear for discrete and approximately evenly spaced radii. 


\begin{figure}[t]
 \centering
    \includegraphics[width=.65\linewidth]{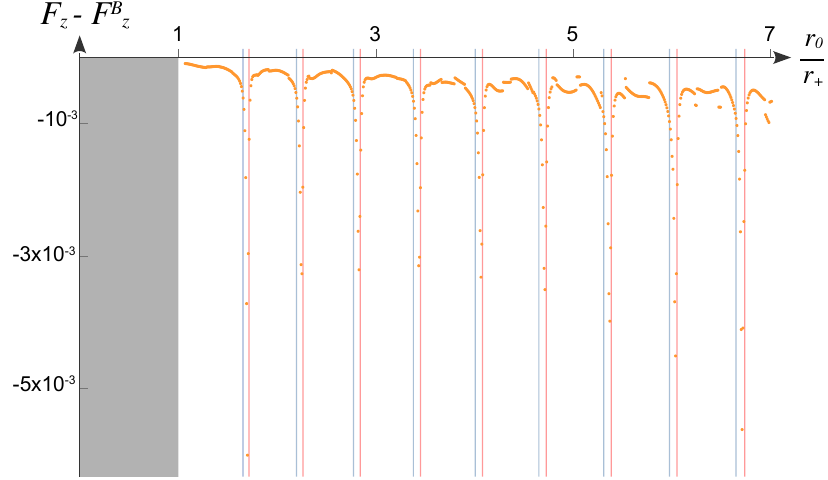}
\caption{Total thrust force obtained as function of the mirror radius $r_0$ normalized with the horizon $r_+$ at $k=0.37$. Here we fix $Q=0.99M$, ${\sf e}=5{\sf m}$, $T=0.07$ and $\mu=-0.01$. The blue and red lines correspond to the zeros of $\mathfrak{Re}[u_\ell(r)]$ and $\mathfrak{Im}[u_\ell(r)]$ respectively for $\ell=0$.}
    \label{fig:force_resonance}
\end{figure}

To further understand such ``resonant'' modes, in Fig.~\ref{fig:force_resonance} we plot the force \eqref{eq:thrust_result} for an isotropic superposition of modes with a fixed value of $k$ as a function of the mirror radius. We see that the resulting peaks coincide almost perfectly with the zeros of the radial function $u_\ell(r)$. This supports the interpretation of the peaks as due to the modes that would resonate on a closed cavity with similar radius.  For a more detailed analysis, see Appendix \ref{appendix:resonance}. 

\section{Discussion}

In this note, we calculated the thrust on a semispherical mirror centered around a charged black hole, which is immersed in a thermal bath made of charged scalar particles. We dubbed the system a ``black hole superradiant rocket''

We obtained the resulting force as a function of the black hole parameters mass and charge, the thermal bath parameters temperature and chemical potential, and the rocket parameter corresponding to the mirror radius. We compared it with the equilibration thrust that we would obtain in the absence of the black hole. 

We found that within the superradiant regime there is a net force acting on the system, originated on the superradiant modes. Furthermore, we identified that by fine-tuning external and rocket parameters, this force can be optimized to maximize its value. Particularly noteworthy are the resonant peaks in the thrust profile, reminiscent of the well-known "black hole bomb" modes \cite{Degollado}. The presence of these peaks can be understood as follows: the resonant modes of the black hole bomb do not disappear as we open a hole in the spherical mirror enclosing the black hole; rather, they shift and are smoothed out, persisting even when the hole becomes a full hemisphere. Leveraging these resonances by adjusting the mirror's radius to coincide with one of these peaks allows for the extraction of maximum thrust.


The results we have obtained represent just one instance of a broader phenomenon: an object near a superradiant black hole experiences a force due to the superradiant modes. We deliberately chose a semispherical mirror to match the horizon's geometry, making it easier to establish a separable system and formulate dynamics through a boundary value problem. However, the mirror's shape could potentially vary to include any arbitrary spherical cap or zone, which would imply adding more $\gamma^m$ matrices in our calculations.

A related but slightly different calculation would involve computing the flux of the energy-momentum tensor on a surface that encloses only the mirror, excluding the black hole. This would yield the force acting on a semispherical sail.

To explore slightly more realistic scenarios, one could attempt to solve the problem for a superradiant Kerr black hole. In such cases, selecting a mirror shape corresponding to a constant radius value within a coordinate system where the scalar equations can be separated is crucial. Additionally, it would be worthwhile to calculate the force originating from a fermionic thermal bath. These considerations are essential steps before applying any of our present results to astrophysical settings, such as estimating the force exerted by a superradiant Kerr black hole on surrounding objects \cite{Arvanitaki,Cardoso,Blas}. 

From an engineering perspective, the force exerted on the mirror depends on how far the black hole is from reaching thermodynamic equilibrium with its environment (which is encoded in the variable $\mu-\mu_{BH}$). To keep the thrust going, a constant supply of charged particles must be fed to the black hole. Since superradiance is a classical phenomenon, the force is not dependent on the value of $\hbar$. This can be compared with the 'black hole starship' proposed in \cite{westmoreland,bolonkin}, which uses the power extracted from Hawking radiation to generate thrust. An interesting question is whether there is a window of parameters in which both phenomena can be combined to obtain a more efficient engine.

\section*{Acknowledgments} 
We are grateful to Diego Correa, Guillermo Silva, Julio Oliva, Octavio Fierro and Shawn Westmoreland for helpful exchange. This work is partially supported by CONICET and UNLP I+D grants X931 and X909.

\newpage

\appendix

\newpage
\section{ENERGY-MOMENTUM TENSOR}\label{appendix:Tmunu}
The Lagrangian of a free complex scalar field is given by:
\begin{equation}
    \mathcal{L}=:(\partial_{\alpha}\hat{\Phi})(\partial^{\alpha}\hat{\Phi}^{\dagger})-{\sf m}^2\hat{\Phi}\hat{\Phi}^{\dagger}:
\end{equation}
From this we can compute its energy-momentum tensor:
\begin{equation}
    \begin{split}
        T_{\mu\nu}&=\frac{\partial\mathcal{L}}{\partial(\partial^{\mu}\hat{\Phi})}\partial_{\nu}\hat{\Phi}
        +\frac{\partial\mathcal{L}}{\partial(\partial^{\mu}\hat{\Phi}^{\dagger})}\partial_{\nu}\hat{\Phi}^{\dagger}-g_{\mu\nu}\mathcal{L}
        \\
        &=\left(\partial_{\mu}\partial'_{\nu} +\partial'_{\mu}\partial_{\nu} -g_{\mu\nu}(\partial'_{\alpha}\partial^{\alpha}-{\sf m}^2)\right)
        \left(:\hat{\Phi}^{\dagger}(\mathbf{x})\hat{\Phi}(\mathbf{x'}):\right)\arrowvert_{\mathbf{x'}=\mathbf{x}}
    \end{split}
\end{equation}
where we write $\partial_{\mu}\hat{\Phi}(\mathbf{x})=\partial_{\mu}'\hat{\Phi}(\mathbf{x'})\arrowvert_{\mathbf{x'}=\mathbf{x}}$ in order to simplify the notation. However, since we are interested in a thermal bath of particles and antiparticles at temperature $T$ and chemical potential $\mu$, then it will be more useful to take the mean value of the energy-momentum tensor:
\begin{equation}
    T_{\mu\nu}=\frac{1}{2} \left(\partial_{\mu}\partial'_{\nu} +\partial'_{\mu}\partial_{\nu} -g_{\mu\nu}(\partial'_{\alpha}\partial^{\alpha}-{\sf m}^2)\right)
   \braket{:\hat{\Phi}^{\dagger}(\mathbf{x})\hat{\Phi}(\mathbf{x'}):}\arrowvert_{\mathbf{x'}=\mathbf{x}}
\end{equation}
the prefactor $1/2$ is to avoid double counting. Finally, we couple the gauge field, 
\begin{equation}
    \left\{
    \begin{aligned}
        \partial_{\mu}\hat{\Phi}^{\dagger}(\mathbf{x})&\longrightarrow\partial_{\mu}\hat{\Phi}^{\dagger}(\mathbf{x})+i{\sf e}A_{\mu}\hat{\Phi}^{\dagger}(\mathbf{x})
        \\
        \partial'_{\mu}\hat{\Phi}(\mathbf{x'})&\longrightarrow\partial'_{\mu}\hat{\Phi}(\mathbf{x'})-i{\sf e}A'_{\mu}\hat{\Phi}(\mathbf{x'})
    \end{aligned}
    \right.
\end{equation}
and obtain the energy-momentum tensor: 
\begin{multline}\label{eq:Tmunu}
    T_{\mu\nu}=\frac{1}{2} \left((\partial_{\mu}+i{\sf e}A_{\mu})(\partial'_{\nu}-i{\sf e}A'_{\nu}) +(\partial'_{\mu}-i{\sf e}A'_{\mu})(\partial_{\nu}+i{\sf e}A_{\nu})\right.
    \\
    \left.-g_{\mu\nu}((\partial'_{\alpha}-i{\sf e}A'_{\alpha})(\partial^{\alpha}+i{\sf e}A^{\alpha})-{\sf m}^2)\right)\braket{:\hat{\Phi}^{\dagger}(\mathbf{x})\hat{\Phi}(\mathbf{x'}):}\arrowvert_{\mathbf{x'}=\mathbf{x}}
\end{multline}

Now, for the calculation of the thrust force in Sec. \ref{sec:thrust_calculation} we only need to compute $T_{rr}$ and $T_{r\theta}$ in the limit $r\rightarrow\infty$. In this limit, the gauge field will be just $A_\mu=A'_\mu=\mu$, then for the components of interest we have: 
\begin{align}
    T_{rr}&=\frac{1}{2} \left(\partial'_{t}\partial_{t}+i{\sf e}\mu(\partial'_{t}-\partial_{t}) 
    +\partial_{r}\partial'_{r} -\frac{1}{r^2}\partial'_{\theta}\partial_{\theta} -\frac{1}{r^2\sin^2(\theta)}\partial'_{\phi}\partial_{\phi} -{\sf m}^2+{\sf e}^2\mu^2\right)\times \nonumber    \\
    &\hspace{9cm}\times
    \braket{:\hat{\Phi}^{\dagger}(\mathbf{x})\hat{\Phi}(\mathbf{x'}):}\arrowvert_{\mathbf{x'}=\mathbf{x}} \label{eq:Trr}  \\
    T_{r\theta}&=\frac{1}{2} \left(\partial_{r}\partial'_{\theta} +\partial'_{r}\partial_{\theta}\right) 
    \braket{:\hat{\Phi}^{\dagger}(\mathbf{x})\hat{\Phi}(\mathbf{x'}):}\arrowvert_{\mathbf{x'}=\mathbf{x}}  \label{eq:Trtheta}
\end{align}
Therefore, in order to obtain these components of the energy-momentum tensor, it only remains to obtain the expected value of the fields 
$\braket{:\hat{\Phi}^{\dagger}(\mathbf{x})\hat{\Phi}(\mathbf{x'}):}$. Since we are computing the energy-momentum tensor at infinity, the field 
at infinity is given by two contributions $\hat\Phi=\hat\Phi^{\sf inc}+\hat\Phi^{\sf scatt}$, the incident and the scattered components at 
infinity. Thus, we can use this contributions of the field to separate the expectation value as:
\begin{multline}
    \braket{:\hat{\Phi}^{\dagger}(\mathbf{x})\hat{\Phi}(\mathbf{x'}):}\arrowvert_{\mathbf{x'}=\mathbf{x}}=\left[\braket{:\hat{\Phi}^{\sf scat\dagger}(\mathbf{x})\hat{\Phi}^{\sf scat}(\mathbf{x'}):} 
    +\braket{:\hat{\Phi}^{\sf scat\dagger}(\mathbf{x})\hat{\Phi}^{\sf inc}(\mathbf{x'}):}\right.
    \\
    \left.+\braket{:\hat{\Phi}^{\sf inc\dagger}(\mathbf{x})\hat{\Phi}^{\sf scat}(\mathbf{x'}):} +\bcancel{\braket{:\hat{\Phi}^{\sf inc\dagger}(\mathbf{x})
    \hat{\Phi}^{\sf inc}(\mathbf{x'}):}}\right]\arrowvert_{\mathbf{x'}=\mathbf{x}}
\end{multline}
where the pure incident field term vanishes when we integrate in all directions (it is the total force without the mirror). In the same way, the momentum energy tensor can be separated by writing:
\begin{equation}\label{eq:Tmunu2}
    \left\{ 
    \begin{aligned}
        T_{rr}&=T^{\sf scat/scat}_{rr}+T^{\sf scat/inc}_{rr}+T^{\sf inc/scat}_{rr}
        \\
        T_{r\theta}&=T^{\sf scat/scat}_{r\theta}+T^{\sf scat/inc}_{r\theta}+T^{\sf inc/scat}_{r\theta}
    \end{aligned}
    \right.
\end{equation}
In order to obtain each of this terms, we  will need the fields expectation value, but first we will compute the mean value between the operators $\hat{b}_{\ell}^m$ and $\hat{c}_{\ell}^m$ which are present in the field definitions \eqref{eq:outer}.

\subsection{Mean values between operators \texorpdfstring{$\hat{b}_{\ell}^m$}{TEXT} and \texorpdfstring{$\hat{c}_{\ell}^m$}{TEXT} }\label{appendix:angular_mean_values}
Starting from the definition of the incident field operator \eqref{eq:SolutionIncident} but using the normalization of the quantum field \eqref{eq:quantum_field}:
\begin{equation}
    \hat{c}_{\ell}^m=\frac{\omega_++{\sf e}\mu}{2\pi^2\sqrt{|\omega_+\omega_-|}}\frac{e^{i\frac{\pi}{2}\ell}}{\sqrt{2\:\omega_+}}\hat{A}_{\vec{k}}Y_{\ell}^{m*}(\check{n})w_{\ell}(r_0)
\end{equation}
and its analogue for antiparticles,
\begin{equation}
    \hat{\tilde{c}}_\ell^m=\frac{\omega_++{\sf e}\mu}{2\pi^2\sqrt{|\omega_+\omega_-|}}\frac{e^{i\frac{\pi}{2}\ell}}{\sqrt{-2\:\omega_-}}\hat{B}_{\vec{k}}Y_\ell^{m*}(\check{n})w_\ell(r_0)
\end{equation}
where we used $|\omega_{\pm}+{\sf e}\mu|=\omega_++{\sf e}\mu$. We can immediately obtain the expectation value of the quadratic operator $\hat{c}_{\ell}^{m\dagger}\hat{c}_{\ell}^m$ over all directions of incidence: 
\begin{equation}
    \begin{split}
        \int d\check{n} \: d\check{n}' \: \braket{\hat{c}_{\ell}^{m\dagger}\hat{c}_{\ell'}^{m'}} =
            \int d\check{n} \: d\check{n}' \: &\frac{e^{i\frac{\pi}{2}(\ell'-\ell)}}{(2\pi^2)^2\sqrt{2\:\omega_+}\sqrt{2\omega_+'}}\frac{(\omega_++{\sf e}\mu)(\omega_+'+{\sf e}\mu)}{\sqrt{|\omega_+\omega_-|}\sqrt{|\omega_+'\omega_-'|}}\times
            \\
            &\times Y_{\ell}^{m}(\check{n})Y_{\ell'}^{m'*}(\check{n}')w_{\ell}^*(r_0)w_{\ell'}(r_0)\braket{\hat{A}_{\vec{k}}^{\dagger}\hat{A}_{\vec{k}'}}
    \end{split}
\end{equation}
Using the Bose-Einstein distribution \eqref{eq:adaggera} and writing the Dirac delta in spherical coordinates,
\begin{equation}
    \begin{split}
        \int d\check{n} \: d\check{n}' \: \braket{\hat{c}_{\ell}^{m\dagger}\hat{c}_{\ell'}^{m'}} = \int &d\check{n} \: d\check{n}' \: \frac{e^{i\frac{\pi}{2}(\ell'-\ell)}}{\pi\sqrt{\omega_+}\sqrt{\omega_+'}}\frac{(\omega_++{\sf e}\mu)(\omega_+'+{\sf e}\mu)}{\sqrt{|\omega_+\omega_-|}\sqrt{|\omega_+'\omega_-'|}}\times
        \\
        &\times Y_{\ell}^{m}(\check{n})Y_{\ell'}^{m'*}(\check{n}')\frac{w_{\ell}^*(r_0)w_{\ell'}(r_0)}{e^{\frac{\omega_+}{T}}-1}\frac{\delta(k-k)}{k^2}\frac{\delta(\theta-\theta')\delta(\phi-\phi')}{\sin(\theta)}
    \end{split}
\end{equation}
The integral is over all the possible directions but not over $k$, then by doing the angular integration it remains: 
\begin{equation}\label{eq:ev_c}
    \int d\check{n} \: d\check{n}' \: \braket{\hat{c}_{\ell}^{m\dagger}\hat{c}_{\ell'}^{m'}} = \frac{(\omega_++{\sf e}\mu)^2}{\pi\:|\omega_+\omega_-|}\frac{|w_{\ell}(r_0)|^2}{\omega_+\left(e^{\frac{\omega_+}{T}}-1\right)} \frac{\delta(k-k')}{k^2} \delta_{\ell\ell'}\delta_{mm'}
\end{equation}
Proceeding in the same way for antiparticles, the expectation value of the quadratic operator is obtained considering all directions of incidence, where the change of sign in the exponential is due to the change in the Bose-Einstein distribution:
\begin{equation}\label{eq:ev_c_anti}
    \int d\check{n} \: d\check{n}' \: \braket{\hat{\tilde{c}}_{\ell}^{m\dagger}\hat{\tilde{c}}_{\ell'}^{m'}} = -\frac{(\omega_++{\sf e}\mu)^2}{\pi\:|\omega_+\omega_-|}\frac{|w_{\ell}(r_0)|^2}{\omega_-\left(e^{-\frac{\omega_-}{T}}-1\right)} \frac{\delta(k-k')}{k^2} \delta_{\ell\ell'}\delta_{mm'}
\end{equation}

Moreover, using the result \eqref{eq:ev_c} and the classical result \eqref{eq:b.enfuncionde.c} we can obtain the mean value between $\hat{c}_{\ell}^m$ and $\hat{b}_{\ell}^m$ over all directions: 
\begin{equation}\label{eq:expectation_value_c^dagger b}
    \begin{split}
        \int d\check{n} \: d\check{n}' \: \braket{\hat{c}_{\ell}^{m\dagger}\hat{b}_{\ell'}^{m'}} & = \int d\check{n} \: d\check{n}' \: \braket{\hat{c}_{\ell}^{m\dagger} \sum_{\ell''=|m'|}^{\infty}\hat{c}_{\ell''}^{m'}O_{\ell'\ell''}}
        =\sum_{\ell''=|m'|} O_{\ell'\ell''} \int d\check{n} \: d\check{n}' \: \braket{\hat{c}_{\ell}^{m\dagger}\hat{c}_{l''}^{m'}}
        \\ 
        & = O_{\ell'\ell} \frac{(\omega_++{\sf e}\mu)^2}{\pi\:|\omega_+\omega_-|}\frac{|w_{\ell}(r_0)|^2}{\omega_+\left(e^{\frac{\omega_+}{T}}-1\right)} \frac{\delta(k-k')}{k^2} \delta_{mm'}
    \end{split}
\end{equation}
Analogously, other useful results are obtained:
\begin{equation}\label{eq:expectation_value_b^dagger c}
    \int d\check{n} \: d\check{n}' \: \braket{\hat{b}_{\ell}^{m\dagger}\hat{c}_{\ell'}^{m'}} = 
    O_{\ell\ell'}^* \frac{(\omega_++{\sf e}\mu)^2}{\pi\:|\omega_+\omega_-|}\frac{|w_{\ell'}(r_0)|^2}{\omega_+\left(e^{\frac{\omega_+}{T}}-1\right)} \frac{\delta(k-k')}{k^2} \delta_{m'm}
\end{equation}
\begin{equation}\label{eq:expectation_value_b^dagger b}
    \int d\check{n} \: d\check{n}' \: \braket{\hat{b}_{\ell}^{m\dagger}\hat{b}_{\ell'}^{m'}} 
    = \sum_{\ell''=|m|}^{\infty} O_{\ell\ell''}^* O_{\ell'\ell''} \frac{(\omega_++{\sf e}\mu)^2}{\pi\:|\omega_+\omega_-|}\frac{|w_{\ell''}(r_0)|^2}{\omega_+\left(e^{\frac{\omega_+}{T}}-1\right)} \frac{\delta(k-k')}{k^2}\delta_{mm'}
\end{equation}

In addition, we obtain the same expressions between the antiparticles operators $\hat{\tilde{c}}_\ell^m$ and $\hat{\tilde{b}}_\ell^m$ by using \eqref{eq:ev_c_anti}. 
\begin{equation}
    \begin{split}
        \int d\check{n} \: d\check{n}' \: \braket{\hat{\tilde{c}}_{\ell}^{m\dagger}\hat{b}_{\ell'}^{m'}} 
        = -O_{\ell'\ell} \frac{(\omega_++{\sf e}\mu)^2}{\pi\:|\omega_+\omega_-|}\frac{|w_{\ell}(r_0)|^2}{\omega_-\left(e^{-\frac{\omega_-}{T}}-1\right)} \frac{\delta(k-k')}{k^2} \delta_{mm'}
    \end{split}
\end{equation}
\begin{equation}
    \int d\check{n} \: d\check{n}' \: \braket{\hat{b}_{\ell}^{m\dagger}\hat{\tilde{c}}_{\ell'}^{m'}} = 
    -O_{\ell\ell'}^* \frac{(\omega_++{\sf e}\mu)^2}{\pi\:|\omega_+\omega_-|}\frac{|w_{\ell'}(r_0)|^2}{\omega_-\left(e^{-\frac{\omega_-}{T}}-1\right)} \frac{\delta(k-k')}{k^2} \delta_{m'm}
\end{equation}
\begin{equation}
    \int d\check{n} \: d\check{n}' \: \braket{\hat{b}_{\ell}^{m\dagger}\hat{b}_{\ell'}^{m'}} 
    = -\sum_{\ell''=|m|}^{\infty} O_{\ell\ell''}^* O_{\ell'\ell''} \frac{(\omega_++{\sf e}\mu)^2}{\pi\:|\omega_+\omega_-|}\frac{|w_{\ell''}(r_0)|^2}{\omega_-\left(e^{-\frac{\omega_-}{T}}-1\right)} \frac{\delta(k-k')}{k^2}\delta_{mm'}
\end{equation}
We then have the needed tools to compute the expectation values of the fields.

\subsection{Expectation values \texorpdfstring{$\braket{:\hat{\Phi}^{\dagger}(\mathbf{x})\hat{\Phi}(\mathbf{x'}):}$}{TEXT}}\label{appendix:fields_expectation_values}
The quantisized version of the classical fields obtained in sec \ref{sec:infty_fields} are:
\begin{equation}\label{eq:field_scat2}
    \begin{split}
        \hat{\Phi}^{\sf scat}(r,\theta,\phi,t)=
        \int_{\infty}^{\infty} d^3k \: \sum_{\ell,m}&\left( e^{-i\:\omega_+ t}\: \hat{b}_{\ell}^{m}\frac{v_{\ell}(r)}{v_{\ell}(r_0)}Y_{\ell}^m(\theta,\phi)\right.
        \\
        &\left.+ e^{-i\:\omega_- t}\: \hat{\tilde{b}}_{\ell}^{m\dagger} \frac{v_{\ell}^*(r)}{v_{\ell}^*(r_0)}Y_{\ell}^{m*}(\theta,\phi)\right)
    \end{split}
\end{equation}
\begin{equation}\label{eq:field_inc2}
    \begin{split}
        \hat{\Phi}^{\sf inc}(r,\theta,\phi,t)=
        \int_{\infty}^{\infty} d^3k \: \sum_{\ell,m}&\left( e^{-i\:\omega_+ t}\:     \hat{c}_{\ell}^{m}\frac{w_{\ell}(r)}{w_{\ell}(r_0)}Y_{\ell}^m(\theta,\phi)\right.
        \\
        &\left.+ e^{-i\:\omega_- t}\: \hat{\tilde{c}}_{\ell}^{m\dagger} \frac{w_{\ell}^*(r)}{w_{\ell}^*(r_0)}Y_{\ell}^{m*}(\theta,\phi)\right)
    \end{split}
\end{equation}

First we obtain $\braket{:\hat{\Phi}^{\sf scat\dagger}(\mathbf{x})\hat{\Phi}^{\sf scat}(\mathbf{x'}):}$, then replacing with \eqref{eq:field_scat2} and expanding the product we have:
\begin{equation}
    \begin{split}
        \langle:\hat{\Phi}^{\sf scat\dagger}(\mathbf{x})\hat{\Phi}^{\sf scat}&(\mathbf{x'}):\rangle= \int d^3k \: d^3k' \:\sum_{\ell,m}\sum_{\ell',m'}\times
        \\
        \times&\left[ e^{i(\omega_+\: t -\omega_+' t')} \frac{v_{\ell}^*(r)v_{\ell'}(r')}{v_{\ell}^*(r_0)v_{\ell'}(r_0)}Y_\ell^{m*}(\theta,\phi)Y_{\ell'}^{m'}(\theta',\phi')\braket{\hat{b}_\ell^{m\dagger}\hat{b}_{\ell'}^{m'}} \right.
        \\
        +& \left. e^{-i(-\omega_- t+\omega_-' t')} \frac{v_{\ell}(r)v_{\ell'}^*(r')}{v_{\ell}(r_0)v_{\ell'}^*(r_0)}Y_\ell^{m}(\theta,\phi)Y_{\ell'}^{m'*}(\theta',\phi')\braket{\hat{\tilde{b}}_{\ell'}^{m'\dagger}\hat{\tilde{b}}_\ell^{m}}\right]
    \end{split}
\end{equation}
where we used that the only nonzero mean value is $\braket{\hat{b}_\ell^{m\dagger}\hat{b}_{\ell'}^{m'}}$ (as discussed in Sec. \ref{appendix:angular_mean_values}). Also, using spherical coordinates we can use the angular integral \eqref{eq:expectation_value_b^dagger b}, obtaining then the mean value of the fields:
\begin{equation}\label{eq:expectation_value_scat/scat}
    \begin{split}
        \braket{:\hat{\Phi}^{\sf scat\dagger}(\mathbf{x})\hat{\Phi}^{\sf scat}(\mathbf{x'}):}& = \int dk \:\frac{(\omega_++{\sf e}\mu)^2k^2}{\pi\:|\omega_+\omega_-|} \sum_{\ell,\ell',m}\sum_{\ell''=|m|}|w_{\ell''}(r_0)|^2
        \times
        \\
        &\times\left[e^{i\:\omega_+\:(t-t')}\frac{v_{\ell}^*(r)v_{\ell'}(r')}{v_{\ell}^*(r_0)v_{\ell'}(r_0)}\frac{Y_\ell^{m*}(\theta,\phi)Y_{\ell'}^{m}(\theta',\phi') O_{\ell\ell''}^* O_{\ell'\ell''}}{\omega_+\left(e^{\frac{\omega_+}{T}}-1\right)} \right.
        \\
        &\quad \left. - e^{i\:\omega_-(t-t')}  \frac{v_{\ell}(r)v_{\ell'}^*(r')}{v_{\ell}(r_0)v_{\ell'}^*(r_0)}\frac{Y_\ell^{m}(\theta,\phi)Y_{\ell'}^{m*}(\theta',\phi')O_{\ell'\ell''}^* O_{\ell\ell''}}{\omega_-\left(e^{-\frac{\omega_-}{T}}-1\right)} \right]
    \end{split}
\end{equation}

In the same way we can obtain the expectation value $\braket{:\hat{\Phi}^{scat\dagger}(\mathbf{x})\hat{\Phi}^{inc}(\mathbf{x'}):}$, then replacing with \eqref{eq:field_scat2} and \eqref{eq:field_inc2} we have:
\begin{equation}
    \begin{split}
        \langle:\hat{\Phi}^{\sf scat\dagger}(\mathbf{x})\hat{\Phi}^{\sf inc}&(\mathbf{x'}):\rangle= \int d^3k \: d^3k' \:\sum_{\ell,m}\sum_{\ell',m'} \times
        \\
        \times&\left[e^{i(\omega_+ t -\omega_+' t')} \frac{v_{\ell}^*(r)w_{\ell'}(r')}{v_{\ell}^*(r_0)w_{\ell'}(r_0)}Y_\ell^{m*}(\theta,\phi)Y_{\ell'}^{m'}(\theta',\phi')\braket{\hat{b}_\ell^{m\dagger}\hat{c}_{\ell'}^{m'}} \right.
        \\
        +& \left. e^{-i(-\omega_- t+\omega_-' t')} \frac{v_{\ell}(r)w_{\ell'}^*(r')}{v_{\ell}(r_0)w_{\ell'}^*(r_0)}Y_\ell^{m}(\theta,\phi)Y_{\ell'}^{m'*}(\theta',\phi')\braket{\hat{\tilde{c}}_{\ell'}^{m'\dagger}\hat{\tilde{b}}_\ell^{m}}\right]
    \end{split}
\end{equation}
Using the angular integrals \eqref{eq:expectation_value_c^dagger b} and \eqref{eq:expectation_value_b^dagger c}, we then obtain the expectation value of the fields:
\begin{equation}\label{eq:expectation_value_scat/inc}
    \begin{split}
        \braket{:\hat{\Phi}^{\sf scat\dagger}(\mathbf{x})\hat{\Phi}^{\sf inc}(\mathbf{x'}):}
        =&\int \:dk \: \frac{(\omega_++{\sf e}\mu)^2k^2}{\pi\:|\omega_+\omega_-|} \:\sum_{\ell,m,\ell'}|w_{\ell'}(r_0)|^2\times
        \\
        & \times\left[ e^{i\:\omega_+(t-t')} \frac{v_{\ell}^*(r)w_{\ell'}(r')}{v_{\ell}^*(r_0)w_{\ell'}(r_0)}\frac{Y_\ell^{m*}(\theta,\phi)Y_{\ell'}^{m}(\theta',\phi')O_{\ell\ell'}^*}{\omega_+\left(e^{\frac{\omega_+}{T}}-1\right)} \right.
        \\
        & \quad\left. - e^{i\:\omega_-(t- t')} \frac{v_{\ell}(r)w_{\ell'}^*(r')}{v_{\ell}(r_0)w_{\ell'}^*(r_0)}\frac{Y_\ell^{m}(\theta,\phi)Y_{\ell'}^{m*}(\theta',\phi')O_{\ell\ell'}}{\omega_-\left(e^{-\frac{\omega_-}{T}}-1\right)} \right]
    \end{split}
\end{equation}

Finally, we proceed in the same way for $\braket{:\hat{\Phi}^{\sf inc\dagger}(\mathbf{x})\hat{\Phi}^{\sf scat}(\mathbf{x'}):}$, so replacing with \eqref{eq:field_scat2} and \eqref{eq:field_inc2} we have:
\begin{equation}
    \begin{split}
        \langle:\hat{\Phi}^{\sf inc\dagger}(\mathbf{x})\hat{\Phi}^{\sf scat}&(\mathbf{x'}):\rangle=
        \int d^3k \: d^3k' \:\sum_{\ell,m}\sum_{\ell',m'}\times
        \\
        \times&\left[e^{i(\omega_+ t -\omega_+' t')} \frac{w_{\ell}^*(r)v_{\ell'}(r')}{w_{\ell}^*(r_0)v_{\ell'}(r_0)}Y_{\ell}^{m*}(\theta,\phi)Y_{\ell'}^{m'}(\theta',\phi')\braket{\hat{c}_{\ell}^{m\dagger}\hat{b}_{\ell'}^{m'}} \right.
        \\
        & \left. + e^{-i(-\omega_- t+\omega_-' t')} \frac{w_{\ell}(r)v_{\ell'}^*(r')}{w_{\ell}(r_0)v_{\ell'}^*(r_0)}Y_\ell^{m}(\theta,\phi)Y_{\ell'}^{m'*}(\theta',\phi')\braket{\hat{\tilde{b}}_{r_0'}^{m'\dagger}\hat{\tilde{c}}_{\ell}^{m}}\right]
    \end{split}
\end{equation}
Using the angular integrals \eqref{eq:expectation_value_c^dagger b} and \eqref{eq:expectation_value_b^dagger c}, we then obtain the expectation value of the fields:
\begin{equation}\label{eq:expectation_value_inc/scat}
    \begin{split}
        \braket{:\hat{\Phi}^{\sf inc\dagger}(\mathbf{x})\hat{\Phi}^{\sf scat}(\mathbf{x'}):}
        =&\int \:dk\: \frac{(\omega_++{\sf e}\mu)^2k^2}{\pi\:|\omega_+\omega_-|} \:\sum_{\ell,m,\ell'}|w_{\ell}(r_0)|^2\frac{}{}\times
        \\
        & \times\left[ e^{i\:\omega_+(t- t')}\frac{w_{\ell}^*(r)v_{\ell'}(r')}{w_{\ell}^*(r_0)v_{\ell'}(r_0)}\frac{Y_\ell^{m*}(\theta,\phi)Y_{\ell'}^{m}(\theta',\phi')O_{\ell'\ell}}{\omega_+\left(e^{\frac{\omega_+}{T}}-1\right)}  \right.
        \\
        & \quad\left. -e^{i\:\omega_-(t- t')} \frac{w_{\ell}(r)v_{\ell'}^*(r')}{w_{\ell}(r_0)v_{\ell'}^*(r_0)}\frac{Y_\ell^{m}(\theta,\phi)Y_{\ell'}^{m*}(\theta',\phi')O_{\ell'\ell}^*}{\omega_-\left(e^{-\frac{\omega_-}{T}}-1\right)} \right]
    \end{split}
\end{equation}

We can now continue with the computation of each term of the energy-momentum tensor in \eqref{eq:Tmunu2}.

\subsection{Computation of \texorpdfstring{$T_{rr}$}{TEXT}}\label{appendix:Trr}
We obtain the component $T_{rr}$  of the energy-momentum tensor separating the different field contributions: 
\begin{equation}
    T_{rr}=T^{\sf scat/scat}_{rr}+T^{\sf scat/inc}_{rr}+T^{\sf inc/scat}_{rr}
\end{equation}
where each of this part satisfies the definition \eqref{eq:Trr}:
\begin{multline}
    T_{rr}=\frac{1}{2} \left(\partial'_{t}\partial_{t}+i{\sf e}\mu(\partial'_{t}-\partial_{t}) +\partial_{r}\partial'_{r} -\frac{1}{r^2}\partial'_{\theta}\partial_{\theta} -\frac{1}{r^2\sin^2(\theta)}\partial'_{\phi}\partial_{\phi} -{\sf m}^2+{\sf e}^2\mu^2\right)\times
    \\
    \times
    \braket{:\hat{\Phi}^{\dagger}(\mathbf{x})\hat{\Phi}(\mathbf{x'}):}\arrowvert_{\mathbf{x'}=\mathbf{x}}
\end{multline}

Now we need to replace with each fields expectation value computed in Sec. \ref{appendix:fields_expectation_values}. Beginning with $T^{\sf scat/scat}_{rr}$, we will use the result \eqref{eq:expectation_value_scat/scat}:
\begin{multline}
    T_{rr}^{\sf scat/scat}=\frac{1}{2} \left(\partial'_{t}\partial_{t}+i{\sf e}\mu(\partial'_t-\partial_t) +\partial_{r}\partial'_{r} -\frac{1}{r^2}\partial'_{\theta}\partial_{\theta} -\frac{1}{r^2\sin^2(\theta)}\partial'_{\phi}\partial_{\phi} -{\sf m}^2+{\sf e}^2\mu^2\right)
    \\
    \times\braket{:\hat{\Phi}^{\sf scat\dagger}(\mathbf{x})\hat{\Phi}^{\sf scat}(\mathbf{x'}):}\arrowvert_{\mathbf{x'}=\mathbf{x}}
\end{multline}
since we are interesting in the limit $r\rightarrow\infty$, the angular derivatives will vanish at leader order (since the angular dependence are only in the spherical harmonics). First we will obtain the terms with time derivatives $\partial_t$: 
\begin{equation}
    \begin{split}
        \partial'_t\partial_t\langle:\hat{\Phi}^{\sf scat\dagger}&(\mathbf{x})\hat{\Phi}^{\sf scat}(\mathbf{x'}):\rangle= \int dk \:\frac{(\omega_++{\sf e}\mu)^2k^2}{\pi\:|\omega_+\omega_-|} \sum_{\ell,\ell',m}\sum_{\ell''=|m|}|w_{\ell''}(r_0)|^2\frac{}{}\times
        \\
        \times&\left[\omega_+^2e^{i\:\omega_+(t-t')} \frac{v_{\ell}^*(r)v_{\ell'}(r')}{v_{\ell}^*(r_0)v_{\ell'}(r_0)}\frac{Y_\ell^{m*}(\theta,\phi)Y_{\ell'}^{m}(\theta',\phi') O_{\ell\ell''}^* O_{\ell'\ell''}}{\omega_+\left(e^{\frac{\omega_+}{T}}-1\right)}  \right.
        \\
        & \left. - \omega_-^2e^{i\:\omega_-(t-t')}  \frac{v_{\ell}(r)v_{\ell'}^*(r')}{v_{\ell}(r_0)v_{\ell'}^*(r_0)}\frac{Y_\ell^{m}(\theta,\phi)Y_{\ell'}^{m*}(\theta',\phi')O_{\ell'\ell''}^* O_{\ell\ell''}}{\omega_-\left(e^{-\frac{\omega_-}{T}}-1\right)} \right]
    \end{split}
\end{equation}
and 
\begin{equation}
    \begin{split}
        i{\sf e}\mu(\partial'_t-\partial_t&)\langle:\hat{\Phi}^{\sf scat\dagger}(\mathbf{x})\hat{\Phi}^{\sf scat}(\mathbf{x'}):\rangle= \int dk \:\frac{(\omega_++{\sf e}\mu)^2k^2}{\pi\:|\omega_+\omega_-|} \sum_{\ell,\ell',m}\sum_{\ell''=|m|}|w_{\ell''}(r_0)|^2\times
        \\
        \times&\left[2{\sf e}\mu\omega_+e^{i\:\omega_+(t-t')} \frac{v_{\ell}^*(r)v_{\ell'}(r')}{v_{\ell}^*(r_0)v_{\ell'}(r_0)}\frac{Y_\ell^{m*}(\theta,\phi)Y_{\ell'}^{m}(\theta',\phi') O_{\ell\ell''}^* O_{\ell'\ell''}}{\omega_+\left(e^{\frac{\omega_+}{T}}-1\right)}  \right.
        \\
        &\left. -2{\sf e}\mu\omega_- e^{i\:\omega_-(t-t')}  \frac{v_{\ell}(r)v_{\ell'}^*(r')}{v_{\ell}(r_0)v_{\ell'}^*(r_0)}\frac{Y_\ell^{m}(\theta,\phi)Y_{\ell'}^{m*}(\theta',\phi')O_{\ell'\ell''}^* O_{\ell\ell''}}{\omega_-\left(e^{-\frac{\omega_-}{T}}-1\right)} \right]
    \end{split}
\end{equation}
Putting these terms together and using the dispersion relation \eqref{eq:SolutionScattededInfinity}, we can write for the energy-momentum tensor:
\begin{equation}\label{eq:T_rr_time}
    T_{rr}^{\sf scat/scat}=\frac{1}{2} \left(k^2 +\partial_{r}\partial'_{r}\right)\braket{:\hat{\Phi}^{\sf scat\dagger}(\mathbf{x})\hat{\Phi}^{\sf scat}(\mathbf{x'}):}\arrowvert_{\mathbf{x'}=\mathbf{x}}
\end{equation}
Now all that remains to be done are the radial derivatives. Since the function $v_\ell(r)$ satisfies the boundary condition at infinity (discussed in Sec. \ref{sec:infty_fields}) given by:
\begin{equation}\label{eq:v(r)2}
    v_\ell(r)\simeq e^{-i\frac{\pi}{2} (\ell+1)}\frac{e^{i(kr-\eta\ln{2kr})}}{kr}
\end{equation}
and whose derivative at leading order is: 
\begin{equation}\label{eq:v_derivative}
    \partial_r v_\ell(r) \simeq  i\:e^{-i\frac{\pi}{2} (\ell+1)}\frac{e^{i(kr-\eta\ln{2kr})}}{r} = ik v_\ell(r)
\end{equation}
Then, at leading order in this limit, we have for the component $T^{\sf scat/scat}_{rr}$:
\begin{equation}\label{eq:T_rr_aux1}
    T_{rr}^{\sf scat/scat}=k^2\braket{:\hat{\Phi}^{\sf scat\dagger}(\mathbf{x})\hat{\Phi}^{\sf scat}(\mathbf{x'}):}\arrowvert_{\mathbf{x'}=\mathbf{x}}
\end{equation}
Now, replacing with \eqref{eq:expectation_value_scat/scat}, using the asymptotic limit \eqref{eq:v(r)2} of $v_\ell(r)$ and taking $\mathbf{x'}=\mathbf{x}$, we obtain:
\begin{equation}
    \begin{split}
        &T_{rr}^{\sf scat/scat}=\int_0^{\infty} dk \: \frac{(\omega_++{\sf e}\mu)^2k^2}{\pi\:|\omega_+\omega_-|}\sum_{\ell,\ell',m}\sum_{\ell''=|m|}\frac{|w_{\ell''}(r_0)|^2}{r^2}\times 
        \\
        & \left[\frac{e^{i\frac{\pi}{2} (\ell-\ell')}O_{\ell\ell''}^* O_{\ell'\ell''}} {\omega_+\left(e^{\frac{\omega_+}{T}}-1\right)} \frac{Y_\ell^{m*}(\theta,\phi)Y_{\ell'}^{m}(\theta,\phi)}{v_{\ell}^*(r_0)v_{\ell'}(r_0)} 
        -\frac{e^{-i\frac{\pi}{2} (\ell-\ell')}O_{\ell'\ell''}^* O_{\ell\ell''}} {\omega_-\left(e^{-\frac{\omega_-}{T}}-1\right)}   \frac{Y_\ell^{m}(\theta,\phi)Y_{\ell'}^{m*}(\theta,\phi)}{v_{\ell}(r_0)v_{\ell'}^*(r_0)} \right]
    \end{split}
\end{equation}
Moreover, noting that it can be exchanged $\ell\leftrightarrow \ell'$ since it is summing over all possible values of $\ell$ and $\ell'$, then we finally have:
\begin{equation}\label{eq:Trr_scat/scat}
    \begin{split}
        T_{rr}^{\sf scat/scat}=\int_0^{\infty} dk \: &\frac{(\omega_++{\sf e}\mu)^2k^2}{\pi\:|\omega_+\omega_-|}\sum_{\ell,\ell',m}\sum_{\ell''=|m|}  \frac{|w_{\ell''}(r_0)|^2}{r^2} e^{i\frac{\pi}{2} (\ell-\ell')}O_{\ell\ell''}^* O_{\ell'\ell''} \frac{Y_\ell^{m*}(\theta,\phi)Y_{\ell'}^{m}(\theta,\phi)}{v_{\ell}^*(r_0)v_{\ell'}(r_0)}\times  
        \\
        & \quad\times\left[\frac{1}{\omega_+\left(e^{\frac{\omega_+}{T}}-1\right)} -\frac{1}{\omega_-\left(e^{-\frac{\omega_-}{T}}-1\right)} \right]
    \end{split}
\end{equation}

We compute now the component $T_{rr}^{\sf scat/inc}$ in the same way by using the definition of $T_{rr}$ in \eqref{eq:Trr} and the expectation value obtained in \eqref{eq:expectation_value_scat/inc}. First, notice that the time dependence is the same than in the previous case with $\sf scat/scat$ fields (since the difference between the fields are in the radial functions), then using \eqref{eq:T_rr_time} we have:
\begin{equation}
    \begin{split}
        T^{\sf scat/inc}_{rr}=& \frac{1}{2} \left(k^2 +\partial_{r}\partial'_{r} \right)
        \braket{:\hat{\Phi}^{\sf scat\dagger}(\mathbf{x})\hat{\Phi}^{\sf inc}(\mathbf{x'}):}\arrowvert_{\mathbf{x'}=\mathbf{x}}
        \\
        =& \int \:dk \: \frac{(\omega_++{\sf e}\mu)^2k^2}{2\pi\:|\omega_+\omega_-|} \:\sum_{\ell,m,\ell'}|w_{\ell'}(r_0)|^2\times
        \\
        &\:\times\left[ \left(k^2+\partial_{r}\partial'_{r}\right)e^{i\:\omega_+(t-t')} \frac{v_{\ell}^*(r)w_{\ell'}(r')}{v_{\ell}^*(r_0)w_{\ell'}(r_0)}\frac{Y_\ell^{m*}(\theta,\phi)Y_{\ell'}^{m}(\theta',\phi')O_{\ell\ell'}^*}{\omega_+\left(e^{\frac{\omega_+}{T}}-1\right)}\right.
        \\
        & \left.\:- \left(k^2+\partial_{r}\partial'_{r}\right)e^{i\:\omega_-(t-t')}\frac{v_{\ell}(r)w_{\ell'}^*(r')}{v_{\ell}(r_0)w_{\ell'}^*(r_0)}\frac{Y_\ell^{m}(\theta,\phi)Y_{\ell'}^{m*}(\theta',\phi')O_{\ell\ell'}}{\omega_-\left(e^{-\frac{\omega_-}{T}}-1\right)} \right]_{\mathbf{x'}=\mathbf{x}}
    \end{split}
\end{equation}
Again, to compute the radial derivatives we use the asymptotic of $v_\ell(r)$ \eqref{eq:v(r)2} but we also need the asymptotic behavior of $w_\ell(r)$ (also discussed in Sec. \ref{sec:infty_fields}):
\begin{equation}\label{eq:w(r)2}
    w_\ell(r)\simeq i\frac{e^{-i(kr-\eta\ln{2kr}+\ell\frac{\pi}{2})}}{2kr}\left((-1)^\ell-e^{i 2(kr-\eta\ln{2kr})}\right)
\end{equation}
and whose derivative at leader order is: 
\begin{equation}\label{eq:w(r)_derivative}
    \partial_r w_\ell(r)\simeq \frac{e^{-i(kr-\eta\ln{2kr}+\ell\frac{\pi}{2})}\left((-1)^\ell+e^{i 2(kr-\eta\ln{2kr})}\right)}{2r}
\end{equation}
Then, replacing in $T^{\sf scat/inc}_{rr}$ with this asymptotic functions behavior and taking $\mathbf{x'}=\mathbf{x}$ we have:
\begin{equation}
    \begin{split}
        T^{\sf scat/inc}_{rr}= \int \:dk \: \frac{(\omega_++{\sf e}\mu)^2k^2}{\pi\:|\omega_+\omega_-|}\sum_{\ell,m,\ell'}\frac{|w_{\ell'}(r_0)|^2}{2r^2} & \left[\frac{e^{i\frac{\pi}{2} (\ell-\ell')}}{v_{\ell}^*(r_0)w_{\ell'}(r_0)}\frac{Y_\ell^{m*}(\theta,\phi)Y_{\ell'}^{m}(\theta,\phi)O_{\ell\ell'}^*}{\omega_+\left(e^{\frac{\omega_+}{T}}-1\right)}  \right.
        \\
        &\left.
        -\frac{e^{-i\frac{\pi}{2} (\ell-\ell')}}{v_{\ell}(r_0)w_{\ell'}^*(r_0)} \frac{Y_\ell^{m}(\theta,\phi)Y_{\ell'}^{m*}(\theta,\phi)O_{\ell\ell'}}{\omega_-\left(e^{-\frac{\omega_-}{T}}-1\right)} 
        \right]
    \end{split}
\end{equation}
Since $\ell$ and $\ell'$ are dummy indices, we can simplify a bit by interchanging $\ell \leftrightarrow \ell'$ in the second line, finally getting:
\begin{equation}\label{eq:Trr_scat/inc}
    \begin{split}
        T^{\sf scat/inc}_{rr}= \int \:dk &\:\frac{(\omega_++{\sf e}\mu)^2}{\pi\:|\omega_+\omega_-|}\frac{k^2}{2r^2}\sum_{\ell,m,\ell'}  \:e^{i\frac{\pi}{2} (\ell-\ell')}Y_\ell^{m*}(\theta,\phi)Y_{\ell'}^{m}(\theta,\phi)\times
        \\
        &\times\left[\frac{w_{\ell'}^*(r_0)}{v_{\ell}^*(r_0)}\frac{O_{\ell\ell'}^*}{\omega_+\left(e^{\frac{\omega_+}{T}}-1\right)}
        -\frac{w_{\ell}(r_0)}{v_{\ell'}(r_0)}\frac{O_{\ell'\ell}}{\omega_-\left(e^{-\frac{\omega_-}{T}}-1\right)} 
        \right]
    \end{split}
\end{equation}

We compute now the component $T_{rr}^{\sf inc/scat}$ in the same way by using the definition of $T_{rr}$ in \eqref{eq:Trr} and the expectation value obtained in \eqref{eq:expectation_value_inc/scat}. 
Again, the time dependence is the same as in the other cases. Then, using \eqref{eq:T_rr_time} we have:
\begin{equation}
    \begin{split}
        T^{\sf inc/scat}_{rr}&= \frac{1}{2} \left(k^2+\partial_{r}\partial'_{r}\right)
        \braket{:\hat{\Phi}^{\sf inc\dagger}(\mathbf{x})\hat{\Phi}^{\sf scat}(\mathbf{x'}):}\arrowvert_{\mathbf{x'}=\mathbf{x}}
        \\
        &= \int \:dk \: \frac{(\omega_++{\sf e}\mu)^2k^2}{2\pi\:|\omega_+\omega_-|} \:\sum_{\ell,m,\ell'}|w_{\ell}(r_0)|^2\times
        \\
        &\quad\:\times\left[ \left(k^2 +\partial_{r}\partial'_{r} \right)e^{i\omega_+(t-t')} \frac{w_{\ell}^*(r)v_{\ell'}(r')}{w_{\ell}^*(r_0)v_{\ell'}(r_0)}\frac{Y_\ell^{m*}(\theta,\phi)Y_{\ell'}^{m}(\theta',\phi')O_{\ell'\ell}}{\omega_+\left(e^{\frac{\omega_+}{T}}-1\right)}\right.
        \\
        & \qquad\left.\: -\left(k^2+\partial_{r}\partial'_{r} \right)e^{i\:\omega_-(t-t')}\frac{w_{\ell}(r)v_{\ell'}^*(r')}{w_{\ell}(r_0)v_{\ell'}^*(r_0)}\frac{Y_\ell^{m}(\theta,\phi)Y_{\ell'}^{m*}(\theta',\phi')O_{\ell'\ell}^*}{\omega_-\left(e^{-\frac{\omega_-}{T}}-1\right)} \right]_{\mathbf{x'}=\mathbf{x}}
    \end{split}
\end{equation}
Replacing with the asymptotic functions behavior and taking $\mathbf{x'}=\mathbf{x}$ we have:
\begin{equation}
    \begin{split}
        T^{\sf inc/scat}_{rr}= \int \:dk \:\frac{(\omega_++{\sf e}\mu)^2k^2}{\pi\:|\omega_+\omega_-|}\sum_{\ell,m,\ell'}\frac{|w_{\ell}(r_0)|^2}{2r^2}&\left[\frac{e^{-i\frac{\pi}{2} (\ell'-\ell)}}{w_{\ell}^*(r_0)v_{\ell'}(r_0)}\frac{Y_\ell^{m*}(\theta,\phi)Y_{\ell'}^{m}(\theta,\phi)O_{\ell'\ell}}{\omega_+\left(e^{\frac{\omega_+}{T}}-1\right)}\right.
        \\
        &\left.\hspace{-0.3cm} -\frac{e^{i\frac{\pi}{2} (\ell'-\ell)}}{w_{\ell}(r_0)v_{\ell'}^*(r_0)}\frac{Y_\ell^{m}(\theta,\phi)Y_{\ell'}^{m*}(\theta,\phi)O_{\ell'\ell}^*}{\omega_-\left(e^{-\frac{\omega_-}{T}}-1\right)}
        \right]
    \end{split}
\end{equation}
Since $\ell$ and $\ell'$ are dummy indices, we can simplify a bit by interchanging $\ell \leftrightarrow \ell'$ in the first line, finally getting:
\begin{equation}
    \begin{split}
        T^{\sf inc/scat}_{rr}= \int \:dk \:&\frac{(\omega_++{\sf e}\mu)^2}{\pi\:|\omega_+\omega_-|}\frac{k^2}{2r^2}\sum_{\ell,m,\ell'}\:e^{-i\frac{\pi}{2} (\ell-\ell')}Y_{\ell}^{m}(\theta,\phi)Y_{\ell'}^{m*}(\theta,\phi)\times
        \\
        &\times\left[\frac{w_{\ell'}(r_0)}{v_{\ell}(r_0)}\frac{O_{\ell\ell'}}{\omega_+\left(e^{\frac{\omega_+}{T}}-1\right)} -\frac{w_{\ell}^*(r_0)}{v_{\ell'}^*(r_0)}\frac{O_{\ell'\ell}^*}{\omega_-\left(e^{-\frac{\omega_-}{T}}-1\right)}
        \right]
    \end{split} 
\end{equation}
Note that comparing with the component $T^{\sf scat/inc}_{rr}$ in \eqref{eq:Trr_scat/inc} we have:
\begin{equation}\label{eq:Trr_inc/scat}
        T^{\sf inc/scat}_{rr}= \left(T^{\sf scat/inc}_{rr}\right)^*
\end{equation}

Lastly, bringing together the three calculated terms in $T_{rr}$ (expressions \ref{eq:Trr_scat/scat}, \ref{eq:Trr_scat/inc} and \ref{eq:Trr_inc/scat}) and multiplying by $r^2$ (as appears in the total thrust force in Sec. \ref{sec:thrust_calculation}) we obtain:
\begin{equation}
    \begin{split}
        r^2T_{rr}=& r^2T_{rr}^{\sf scat/scat}+2r^2\mathfrak{Re}\left[T_{rr}^{\sf scat/inc}\right]
        \\
        =& \int_0^{\infty} dk \: \frac{(\omega_++{\sf e}\mu)^2k^2}{\pi\:|\omega_+\omega_-|}\sum_{\ell,\ell',m} \left\{ \sum_{\ell''=|m|} |w_{\ell''}(r_0)|^2\: e^{i\frac{\pi}{2} (\ell-\ell')} \frac{Y_\ell^{m*}(\theta,\phi)Y_{\ell'}^{m}(\theta,\phi)}{v_{\ell}^*(r_0)v_{\ell'}(r_0)}\times\right.  
        \\
        & \qquad\qquad\times O_{\ell\ell''}^{m*} O^m_{\ell'\ell''}\left(\frac{1}{\omega_+\left(e^{\frac{\omega_+}{T}}-1\right)} -\frac{1} {\omega_-\left(e^{-\frac{\omega_-}{T}}-1\right)} \right)
        \\
        &\quad\qquad+2\:\mathfrak{Re}\left[\frac{1}{2}e^{i\frac{\pi}{2} (\ell-\ell')}Y_\ell^{m*}(\theta,\phi)Y_{\ell'}^{m}(\theta,\phi)\frac{w_{\ell'}^*(r_0)}{v_{\ell}^*(r_0)}\frac{O_{\ell\ell'}^{m*}}{\omega_+\left(e^{\frac{\omega_+}{T}}-1\right)}\right.
        \\
        &\quad\qquad\qquad\left.\left. -\frac{1}{2}e^{i\frac{\pi}{2} (\ell-\ell')}Y_\ell^{m*}(\theta,\phi)Y_{\ell'}^{m}(\theta,\phi)\frac{w_{\ell}(r_0)}{v_{\ell'}(r_0)}\frac{O^m_{\ell'\ell}}{\omega_-\left(e^{-\frac{\omega_-}{T}}-1\right)} 
        \right]\right\}
    \end{split}
\end{equation}
If we now interchange $\ell\leftrightarrow\ell'$ in the last line, replace with the conjugate and regrouping the remaining terms, we can finally write:
\begin{multline}\label{eq:T_rr2}
    r^2T_{rr}= \int_0^{\infty} dk \: \frac{(\omega_++{\sf e}\mu)^2k^2}{\pi\:|\omega_+\omega_-|}\left(\frac{1}{\omega_+\left(e^{\frac{\omega_+}{T}}-1\right)} -\frac{1} {\omega_-\left(e^{-\frac{\omega_-}{T}}-1\right)} \right)\times
    \\
    \times\sum_{\ell,\ell',m}  \mathcal{U}_{\ell\ell'}Y_\ell^{m*}(\theta,\phi)Y_{\ell'}^{m}(\theta,\phi)
\end{multline}
where we defined
\begin{equation}\label{eq:N_ll'2}
    \mathcal{U}_{\ell\ell'}^m=\sum_{\ell''=|m|} \: e^{i\frac{\pi}{2} (\ell-\ell')} \frac{|w_{\ell''}(r_0)|^2}{v_{\ell}^*(r_0)v_{\ell'}(r_0)} O_{\ell\ell''}^{m*} O^m_{\ell'\ell''}+\mathfrak{Re}\left[e^{i\frac{\pi}{2} (\ell-\ell')}\frac{w_{\ell}(r_0)}{v_{\ell'}(r_0)}O^m_{\ell'\ell}\right]
\end{equation}

\subsection{Computation of \texorpdfstring{$T_{r\theta}$}{TEXT}}
We now compute the other energy-momentum tensor component of interest for calculating the thrust force in Sec. \ref{sec:thrust_calculation}, $rT_{r\theta}$. We use its definition \eqref{eq:Trtheta}: 
\begin{equation}
    T_{r\theta}=\frac{1}{2} \left(\partial_{r}\partial'_{\theta} +\partial'_{r}\partial_{\theta}\right) 
    \braket{:\hat{\Phi}^{\dagger}(\mathbf{x})\hat{\Phi}(\mathbf{x'}):}\arrowvert_{\mathbf{x'}=\mathbf{x}}
\end{equation}
and we calculate this component using each expectation value of the field obtained in Appendix \ref{appendix:fields_expectation_values}. Beginning with the component $T^{\sf scat/scat}_{r\theta}$, we use  \eqref{eq:expectation_value_scat/scat}:
\begin{equation}
    \begin{split}
        rT^{\sf scat/scat}_{r\theta}=&\frac{1}{2}r \left(\partial_{r}\partial'_{\theta} +\partial'_{r}\partial_{\theta}\right) 
        \braket{:\hat{\Phi}^{\sf scat\dagger}(\mathbf{x})\hat{\Phi}^{\sf scat}(\mathbf{x'}):}\arrowvert_{\mathbf{x'}=\mathbf{x}}
        \\
        =&\int dk \:\frac{(\omega_++{\sf e}\mu)^2k^2r}{2\pi\:|\omega_+\omega_-|}\sum_{\ell,\ell',m}\sum_{\ell''=|m|}|w_{\ell''}(r_0)|^2\times
        \\
        &\times\left[e^{i\:\omega_+(t-t')} \frac{\partial_{r}v_{\ell}^*(r)v_{\ell'}(r')}{v_{\ell}^*(r_0)v_{\ell'}(r_0)}\frac{Y_\ell^{m*}(\theta,\phi)\partial'_{\theta}Y_{\ell'}^{m}(\theta',\phi') O_{\ell\ell''}^* O_{\ell'\ell''}}{\omega_+\left(e^{\frac{\omega_+}{T}}-1\right)} \right.
        \\
        &\qquad  - e^{i\:\omega_-(t-t')} \frac{\partial_{r}v_{\ell}(r)v_{\ell'}^*(r')}{v_{\ell}(r_0)v_{\ell'}^*(r_0)}\frac{Y_\ell^{m}(\theta,\phi)\partial'_{\theta}Y_{\ell'}^{m*}(\theta',\phi')O_{\ell'\ell''}^* O_{\ell\ell''}}{\omega_-\left(e^{-\frac{\omega_-}{T}}-1\right)}
        \\
        &\qquad+ e^{i\:\omega_+(t-t')} \frac{v_{\ell}^*(r)\partial'_{r}v_{\ell'}(r')}{v_{\ell}^*(r_0)v_{\ell'}(r_0)}\frac{\partial_{\theta}Y_\ell^{m*}(\theta,\phi)Y_{\ell'}^{m}(\theta',\phi') O_{\ell\ell''}^*O_{\ell'\ell''}}{\omega_+\left(e^{\frac{\omega_+}{T}}-1\right)}  
        \\
        &\qquad \left. - e^{i\:\omega_-(t-t')} \frac{v_{\ell}(r)\partial'_{r}v_{\ell'}^*(r')}{v_{\ell}(r_0)v_{\ell'}^*(r_0)}\frac{\partial_{\theta}Y_\ell^{m}(\theta,\phi)Y_{\ell'}^{m*}(\theta',\phi')O_{\ell'\ell''}^* O_{\ell\ell''}}{\omega_-\left(e^{-\frac{\omega_-}{T}}-1\right)} \right]_{\mathbf{x'}=\mathbf{x}} 
    \end{split}
\end{equation}
Using the asymptotic form of $v_\ell(r)$ in \eqref{eq:v(r)2} and taking $\mathbf{x'}=\mathbf{x}$ we have:
\begin{equation}\label{eq:Trt_scat/scat}
    \begin{split}
        rT^{\sf scat/scat}_{r\theta}&=\int dk \: \frac{(\omega_++{\sf e}\mu)^2}{2\pi\:|\omega_+\omega_-|}\frac{ik}{r} \sum_{\ell,\ell',m}\sum_{\ell''=|m|}|w_{\ell''}(r_0)|^2\times
        \\
        &\times\left[e^{i\frac{\pi}{2} (\ell-\ell')} \frac{O_{\ell\ell''}^* O_{\ell'\ell''}}{v_{\ell}^*(r_0)v_{\ell'}(r_0)} 
        \left(\frac{Y_{\ell'}^{m}(\theta,\phi)\partial_{\theta}Y_\ell^{m*}(\theta,\phi) -Y_\ell^{m*}(\theta,\phi)\partial_{\theta}Y_{\ell'}^{m}(\theta,\phi)}{\omega_+\left(e^{\frac{\omega_+}{T}}-1\right)}\right)\right.
        \\
        &  - \left.e^{-i\frac{\pi}{2} (\ell-\ell')} \frac{O_{\ell'\ell''}^* O_{\ell\ell''}}{v_{\ell}(r_0)v_{\ell'}^*(r_0)} 
        \left(\frac{Y_\ell^{m}(\theta,\phi)\partial_{\theta}Y_{\ell'}^{m*}(\theta,\phi)-Y_{\ell'}^{m*}(\theta,\phi)\partial_{\theta}Y_\ell^{m}(\theta,\phi)}{\omega_-\left(e^{-\frac{\omega_-}{T}}-1\right)}\right)\right]
    \end{split}
\end{equation}
Note that this component is $\mathcal{O}(r^{-1})$.

We obtain in the same way the component $rT^{\sf scat/inc}_{r\theta}$ by using \eqref{eq:expectation_value_scat/inc}: 
\begin{equation}
    \begin{split}
        rT^{\sf scat/inc}_{r\theta}=&\frac{1}{2}r \left(\partial_{r}\partial'_{\theta} +\partial'_{r}\partial_{\theta}\right) 
        \braket{:\hat{\Phi}^{\sf scat\dagger}(\mathbf{x})\hat{\Phi}^{\sf inc}(\mathbf{x'}):}\arrowvert_{\mathbf{x'}=\mathbf{x}}
        \\
        =&
        \int \:dk \: \frac{(\omega_++{\sf e}\mu)^2k^2r}{2\pi\:|\omega_+\omega_-|}\sum_{\ell,m,\ell'}|w_{\ell'}(r_0)|^2\times 
        \\
        &\times\left[ e^{i\:\omega_+(t-t')}\frac{\partial_{r}v_{\ell}^*(r)w_{\ell'}(r')}{v_{\ell}^*(r_0)w_{\ell'}(r_0)}\frac{Y_\ell^{m*}(\theta,\phi)\partial'_{\theta}Y_{\ell'}^{m}(\theta',\phi')O_{\ell\ell'}^*}{\omega_+\left(e^{\frac{\omega_+}{T}}-1\right)}  \right.
        \\
        &\qquad + e^{i\:\omega_+(t-t')}\frac{v_{\ell}^*(r)\partial'_{r}w_{\ell'}(r')}{v_{\ell}^*(r_0)w_{\ell'}(r_0)}\frac{\partial_{\theta}Y_\ell^{m*}(\theta,\phi)Y_{\ell'}^{m}(\theta',\phi')O_{\ell\ell'}^*}{\omega_+\left(e^{\frac{\omega_+}{T}}-1\right)}
        \\
        &\qquad - e^{i\:\omega_-(t- t')} \frac{\partial_{r}v_{\ell}(r)w_{\ell'}^*(r')}{v_{\ell}(R)w_{\ell'}^*(r_0)}\frac{Y_\ell^{m}(\theta,\phi)\partial'_{\theta}Y_{\ell'}^{m*}(\theta',\phi')O_{\ell\ell'}}{\omega_-\left(e^{-\frac{\omega_-}{T}}-1\right)}
        \\
        & \qquad\left. - e^{i\:\omega_-(t- t')} \frac{v_{\ell}(r)\partial'_{r}w_{\ell'}^*(r')}{v_{\ell}(r_0)w_{\ell'}^*(r_0)}\frac{\partial_{\theta}Y_\ell^{m}(\theta,\phi)Y_{\ell'}^{m*}(\theta',\phi')O_{\ell\ell'}}{\omega_-\left(e^{-\frac{\omega_-}{T}}-1\right)}\right]_{\mathbf{x'}=\mathbf{x}}
    \end{split}
\end{equation}
Using the asymptotic form of $v_\ell(r)$ and $w_\ell(r)$ in \eqref{eq:v(r)2} and \eqref{eq:w(r)2} respectively, and taking $\mathbf{x'}=\mathbf{x}$ we have:
\begin{equation}\label{eq:Trt_scat/inc}
    \begin{split}
        rT^{\sf scat/inc}_{r\theta} =&
        \int \:dk \frac{(\omega_++{\sf e}\mu)^2k^2r}{4\pi\:|\omega_+\omega_-|}\frac{k}{r}\sum_{\ell,m,\ell'}|w_{\ell'}(r_0)|^2\times 
        \\
        &\times\left[ \frac{e^{i\frac{\pi}{2} (\ell-\ell'+1)}\left((-1)^{\ell'}e^{-i2kr}-1\right)}{v_{\ell}^*(r_0)w_{\ell'}(r_0)}\frac{Y_\ell^{m*}(\theta,\phi)\partial_{\theta}Y_{\ell'}^{m}(\theta,\phi)O_{\ell\ell'}^*}{\omega_+\left(e^{\frac{\omega_+}{T}}-1\right)}  \right.
        \\
        & \quad+\frac{e^{i\frac{\pi}{2}(\ell-\ell'+1)}\left((-1)^{\ell'}e^{-i2kr}+1\right)}{v_{\ell}^*(r_0)w_{\ell'}(r_0)}\frac{\partial_{\theta}Y_\ell^{m*}(\theta,\phi)Y_{\ell'}^{m}(\theta,\phi)O_{\ell\ell'}^*}{\omega_+\left(e^{\frac{\omega_+}{T}}-1\right)}
        \\
        &\quad - \frac{e^{-i\frac{\pi}{2} (\ell-\ell'+1)}\left((-1)^{\ell'}e^{i2kr}-1\right)}{v_{\ell}(r_0)w_{\ell'}^*(r_0)}\frac{Y_\ell^{m}(\theta,\phi)\partial_{\theta}Y_{\ell'}^{m*}(\theta,\phi)O_{\ell\ell'}}{\omega_-\left(e^{-\frac{\omega_-}{T}}-1\right)}
        \\
        & \quad\left. - \frac{e^{-i\frac{\pi}{2} (\ell-\ell'+1)}\left((-1)^{\ell'}e^{i2kr}+1\right)}{v_{\ell}(r_0)w_{\ell'}^*(r_0)}\frac{\partial_{\theta}Y_\ell^{m}(\theta,\phi)Y_{\ell'}^{m*}(\theta,\phi)O_{\ell\ell'}}{\omega_-\left(e^{-\frac{\omega_-}{T}}-1\right)} \right]
    \end{split}
\end{equation}
Also, note that this component is $\mathcal{O}(r^{-1})$.

Finally we obtain the component $T^{\sf inc/scat}_{r\theta}$ by using \eqref{eq:expectation_value_inc/scat}: 
\begin{equation}
    \begin{split}
        rT^{\sf inc/scat}_{r\theta}=&\frac{1}{2}r \left(\partial_{r}\partial'_{\theta} +\partial'_{r}\partial_{\theta}\right) 
        \braket{:\hat{\Phi}^{\sf inc\dagger}(\mathbf{x})\hat{\Phi}^{\sf scat}(\mathbf{x'}):}\arrowvert_{\mathbf{x'}=\mathbf{x}}
        \\
        =&
        \int \:dk \: \frac{(\omega_++{\sf e}\mu)^2k^2r}{2\pi\:|\omega_+\omega_-|}\sum_{\ell,m,\ell'}|w_{\ell}(r_0)|^2\times 
        \\
        &\times\left[ e^{i\:\omega_+(t- t')}\frac{\partial_{r}w_{\ell}^*(r)v_{\ell'}(r')}{w_{\ell}^*(r_0)v_{\ell'}(r_0)}\frac{Y_\ell^{m*}(\theta,\phi)\partial'_{\theta}Y_{\ell'}^{m}(\theta',\phi')O_{\ell'\ell}}{\omega_+\left(e^{\frac{\omega_+}{T}}-1\right)} \right.
        \\
        &\qquad +e^{i\:\omega_+(t- t')}\frac{w_{\ell}^*(r)\partial'_{r}v_{\ell'}(r')}{w_{\ell}^*(r_0)v_{\ell'}(r_0)}\frac{\partial_{\theta}Y_\ell^{m*}(\theta,\phi)Y_{\ell'}^{m}(\theta',\phi')O_{\ell'\ell}}{\omega_+\left(e^{\frac{\omega_+}{T}}-1\right)}
        \\
        &\qquad - e^{-i\:\omega_+(t- t')} \frac{\partial_{r}w_{\ell}(r)v_{\ell'}^*(r')}{w_{\ell}(r_0)v_{\ell'}^*(r_0)}\frac{Y_\ell^{m}(\theta,\phi)\partial'_{\theta}Y_{\ell'}^{m*}(\theta',\phi')O_{\ell'\ell}^*}{\omega_-\left(e^{-\frac{\omega_-}{T}}-1\right)}
        \\
        & \left.\qquad - e^{-i\:\omega_+(t- t')} \frac{w_{\ell}(r)\partial'_{r}v_{\ell'}^*(r')}{w_{\ell}(r_0)v_{\ell'}^*(r_0)}\frac{\partial_{\theta}Y_\ell^{m}(\theta,\phi)Y_{\ell'}^{m*}(\theta',\phi')O_{\ell'\ell}^*}{\omega_-\left(e^{-\frac{\omega_-}{T}}-1\right)} \right]_{\mathbf{x'}=\mathbf{x}}
    \end{split}
\end{equation}

Using the asymptotic form of $v_\ell(r)$ and $w_\ell(r)$ in \eqref{eq:v(r)2} and \eqref{eq:w(r)2} respectively, and taking $\mathbf{x'}=\mathbf{x}$ we have:
\begin{equation}\label{eq:Trt_inc/scat}
    \begin{split}
        rT^{\sf inc/scat}_{r\theta}
        =&
        \int \:dk \:\frac{(\omega_++{\sf e}\mu)^2}{4\pi\:|\omega_+\omega_-|}\frac{k}{r}\sum_{\ell,m,\ell'}|w_{\ell}(r_0)|^2\times 
        \\
        &\times \left[e^{-i\frac{\pi}{2} (\ell'-\ell+1)}\frac{\left((-1)^\ell e^{i2kr}+1\right)}{w_{\ell}^*(r_0)v_{\ell'}(r_0)}\frac{Y_\ell^{m*}(\theta,\phi)\partial_{\theta}Y_{\ell'}^{m}(\theta,\phi)O_{\ell'\ell}}{\omega_+\left(e^{\frac{\omega_+}{T}}-1\right)}  \right.
        \\
        &\qquad+e^{-i\frac{\pi}{2} (\ell'-\ell+1)}\frac{\left((-1)^\ell e^{i2kr}-1\right)}{w_{\ell}^*(r_0)v_{\ell'}(r_0)}\frac{\partial_{\theta}Y_\ell^{m*}(\theta,\phi)Y_{\ell'}^{m}(\theta,\phi)O_{\ell'\ell}}{\omega_+\left(e^{\frac{\omega_+}{T}}-1\right)}
        \\
        &\qquad - e^{i\frac{\pi}{2} (\ell'-\ell+1)} \frac{\left((-1)^\ell e^{-2ikr}+1\right)}{w_{\ell}(r_0)v_{\ell'}^*(r_0)}\frac{Y_\ell^{m}(\theta,\phi)\partial_{\theta}Y_{\ell'}^{m*}(\theta,\phi)O_{\ell'\ell}^*}{\omega_-\left(e^{-\frac{\omega_-}{T}}-1\right)}
        \\
        & \left.\qquad -e^{i\frac{\pi}{2} (\ell'-\ell+1)} \frac{\left((-1)^\ell e^{-i2kr}-1\right)}{w_{\ell}(r_0)v_{\ell'}^*(r_0)}\frac{\partial_{\theta}Y_\ell^{m}(\theta,\phi)Y_{\ell'}^{m*}(\theta,\phi)O_{\ell'\ell}^*}{\omega_-\left(e^{-\frac{\omega_-}{T}}-1\right)} \right]
    \end{split}
\end{equation}
Again, note that this component is $\mathcal{O}(r^{-1})$.

In conclusion, bringing together the three components of $T_{r\theta}$ we see that this component is $\mathcal{O}(r^{-1})$. On the other hand, the other term in the force is $r^2T_{rr}$ (which we already computed in Sec. \ref{appendix:Trr}) and it is $\mathcal{O}(r^{0})$. Therefore, at leading order when $r\rightarrow\infty$, the component $rT_{r\theta}$ vanishes.

\newpage
\section{NUMERICAL COMPLEMENT}

\subsection{\texorpdfstring{$\gamma^m$}{TEXT} analysis}\label{appendix:gamma}

\begin{figure}[t] 
\centering
\includegraphics[width=0.85\linewidth]{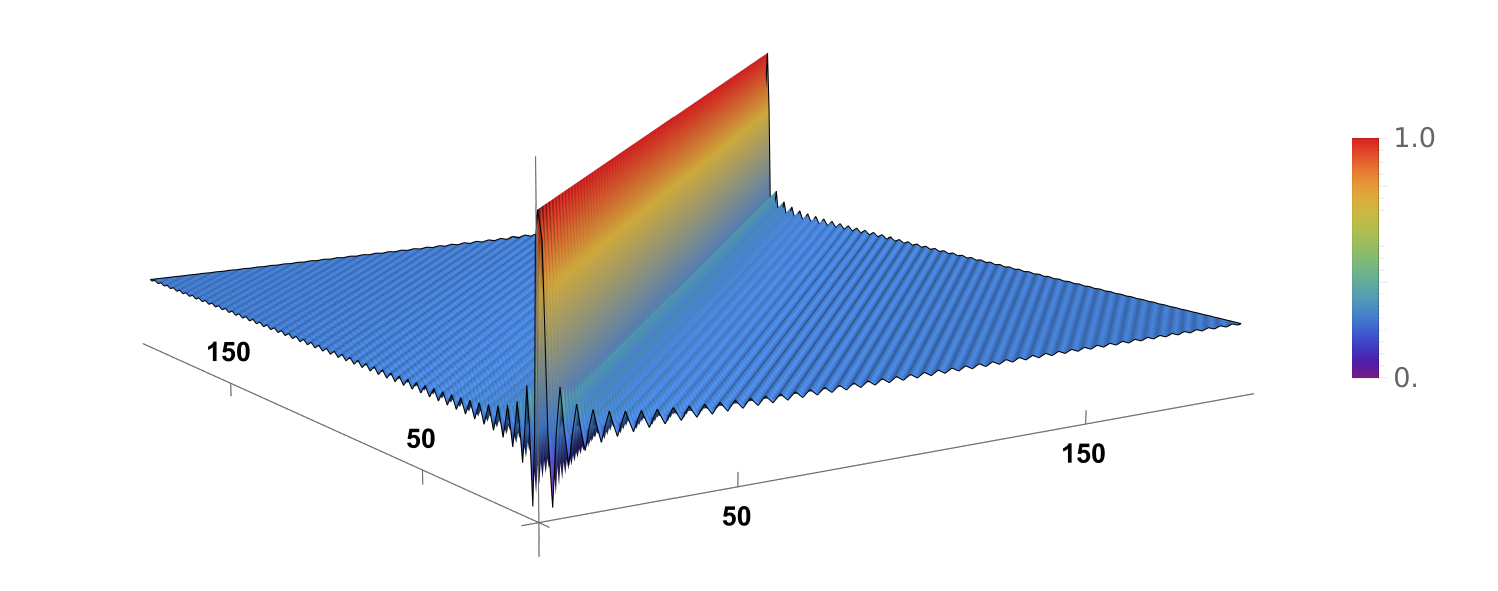}
\put(-260,123){$\gamma^m_{\ell\ell'}$}
\put(-60,50){$\ell'$}
\put(-370,65){$\ell$}
\caption{Matrix of coefficients $\gamma^m_{\ell\ell'}$ with $m=0$.}
\label{fig:gamma}
\end{figure}

The results obtained in Secs. \ref{sec:numerical_rocket} and \ref{sec:numerical_force} depend on $m$ matrices of coefficients $\gamma^m_{\ell\ell'}$ (defined in \eqref{eq:gamma}) and which do not depend on the system parameters. As can be seen in fig \ref{fig:gamma}, these matrices have their highest value on the diagonal $\gamma^m_{\ell\ell}=1$ and oscillate around zero as we move away from the diagonal. This behavior will allows us to cut down the matrix and still grasp the physics of it. 

\subsection{Numerical cutoff}\label{appendix:cut_off}

\begin{figure}[t]
    \centering
    \begin{subfigure}{.47\textwidth}
        \centering
        \includegraphics[width=\linewidth]{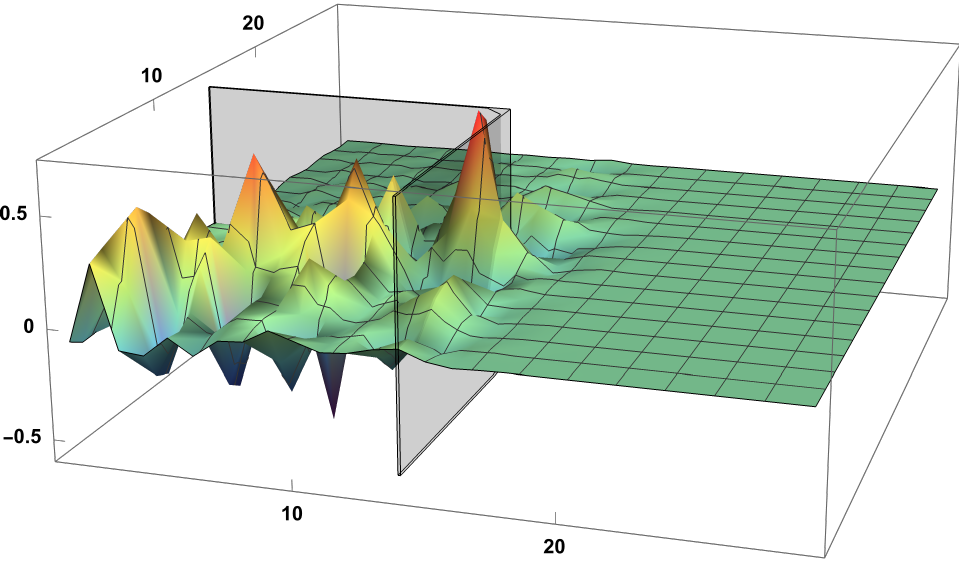}
        \subcaption*{$\mathfrak{Re}\left[O^m_{\ell\ell'}+\delta_{\ell\ell'}\right]$}
    \end{subfigure}%
    \hspace{.06\textwidth}%
    \begin{subfigure}{.47\textwidth}
        \centering
        \includegraphics[width=\linewidth]{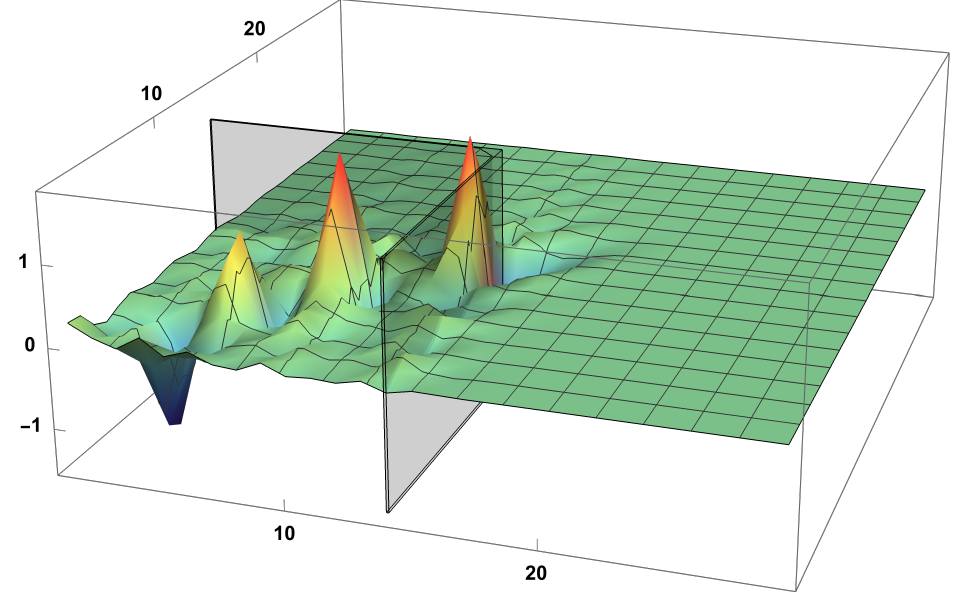}
        \subcaption*{$\mathfrak{Im}\left[O^m_{\ell\ell'}+\delta_{\ell\ell'}\right]$}
    \end{subfigure}
    \caption{Elements of $O^m_{\ell\ell'}+\delta_{\ell\ell'}$ with $m=0$ and $k=0.3$, we mark the value $\ell_{\sf cutoff}=14$ (in this case) where a peak is predicted.}
    \label{fig:U}
\end{figure}

In order to obtain the numerical results we need to defined a cutoff in the sums over $\ell$ (cutting $\ell$ we have cut the sum over $m$ as well since $\ell>|m|$). To do so, first we notice that the result for the matching in Sec. \ref{sec:mirror} depends on the radial functions evaluated at the mirror. Moreover, the result depends indirectly on the inverse of the derivative\footnote{Here we use the notation $R_\ell(r)$ to refer to all the radial solutions $u_\ell(r)$, $v_\ell(r)$ and $w_\ell(r)$.} $R'_\ell(r_{\sf m})$ (these are inside the definition of $H_\ell$ which is inside of $M^m$ defined in \eqref{eq:M}; for our results we need the inverse $(M^m)^{-1}$). Then, analyzing the radial equation \eqref{eq:EOMradial}:
\begin{equation}
    \frac{1}{r^2}\partial_{r}\left(r^2f(r)\partial_{r}R_{\ell}(r)\right)
    +\left(\frac{\left(\omega+{\sf e}\:h(r)\right)^2}{f(r)} -\frac{\ell(\ell+1)}{r^2} -{\sf m}^2
    \right)R_{\ell}(r)=0
\end{equation}
If we define
\begin{equation}
    y_{\ell}(r)=r^2 f(r) R'_{\ell}(r)
\end{equation}
we have:
\begin{align}
    &y'_{\ell}(r)=-{r^2}\left(\frac{(\omega+eh(r))^2}{f(r)}-\frac{\ell(\ell+1)}{r^2}-{\sf m}^2\right)R_{\ell}(r)
    \\
    & R'_{\ell}(r)=\frac {y_{\ell}(r)}{r^2 f(r)}
\end{align}
Note that the parenthesis $\left(\frac{(\omega+eh(r))^2}{f(r)}-\frac{\ell(\ell+1)}{r^2}-{\sf m}^2\right)$ vanishes at a certain value of $\ell$ (for $k$ and $r$ fixed), therefore the derivative $y'_{\ell}(r)$ is zero and we have a minimum of $R'_{\ell}(r)$. Then, for this value of $\ell$ we have a maximum of $(R'_{\ell}(r))^{-1}$, so taking $r=r_{\sf m}$ we define $\ell_{\sf cutoff}$ as:
\begin{equation}\label{eq:l_max}
    \frac{(\omega+{\sf e}h(r_0))^2}{f(r_0)}-\frac{\ell_{\sf cutoff}(\ell_{\sf cutoff}+1)}{r_0^2}-{\sf m}^2=0
\end{equation}
which depends on the value of $k$. This behavior can be seen in Fig. \ref{fig:U}, where the predicted peak is near the $\ell_{\sf cutoff}$. Then, to ensure that we are having a good approximation in the numerical analysis we take $\ell_{\sf max}=\ell_{\sf cutoff}+10$. From Fig. \ref{fig:U} we also see that for $\ell>\ell_{\sf max}$ and far away from the diagonal this does not go to zero, which means that the matching will be better if we take more $\ell$. However, for the thrust force \eqref{eq:thrust_result} we only need the values near the diagonal which we see that go to zero for $\ell>\ell_{\sf max}$. Then, this value is a good cutoff to calculate the thrust force we are interested in.

\subsection{Matching at the mirror}\label{appendix:matching}

\begin{figure}[t]
    \centering
\includegraphics[width=0.47\linewidth]{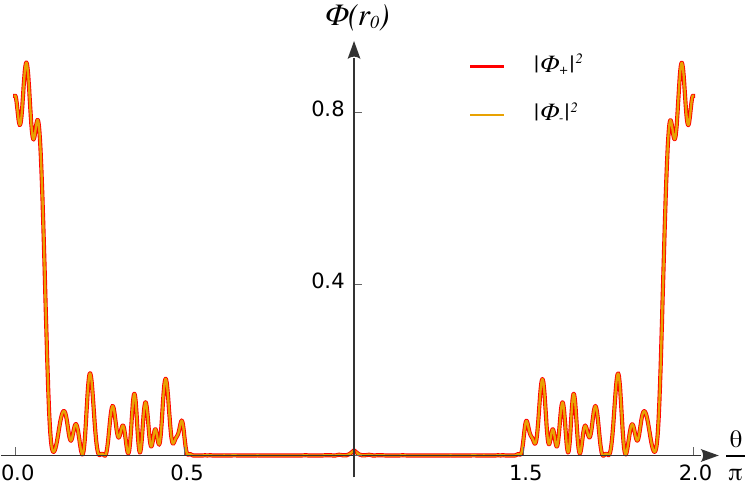}
\hfill
\includegraphics[width=0.47\linewidth]{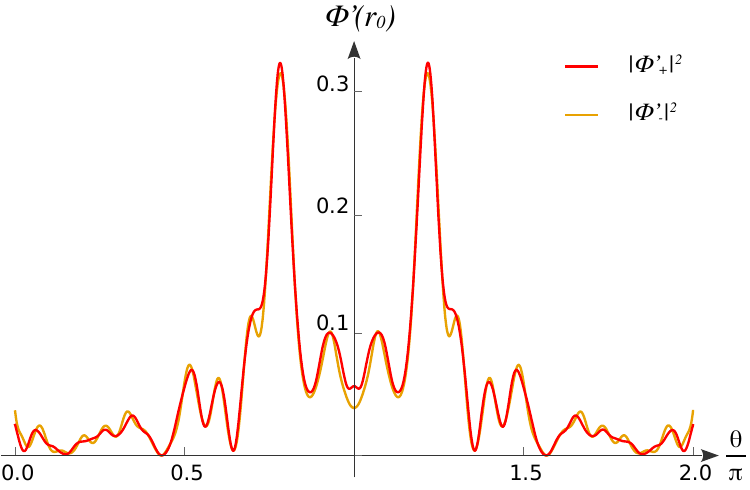}
\caption{Matching conditions of the obtained fields with $k=0.6$ at the mirror radius $r=r_{\sf m}=5\:r_+$ and varying the angle $\theta$. 
Left: density particle of the fields $\Phi^+$ and $\Phi^-$ at the mirror radius in the whole sphere $0\leq\theta\leq2\pi$.
Right: radial derivatives at the mirror radius $\left.\partial_r\Phi^+(r)\right|_{r=r_{\sf m}}$ and $\left.\partial_r\Phi^-(r)\right|_{r=r_{\sf m}}$ on the upper hemisphere $-\frac{\pi}{2}\leq\theta\leq\frac{\pi}{2}$.
}
    \label{fig:Matching}
\end{figure}

In Sec. \ref{sec:numerical_rocket} we show the obtained scalar field in the whole space. In Fig. \ref{fig:Matching} we show how this field fulfills the 3 matching conditions at the mirror radius studied in Sec. \ref{sec:mirror} by using the cutoff for $\ell$ discussed in the previous section. First, we see the fields $\Phi^+$ and $\Phi^-$ at $r=r_{\sf m}$ where we notice that both curves coincide in the whole sphere and are zero at the mirror position, so continuity and the perfect mirror condition is satisfied. Second, we see the radial derivatives of the fields where we notice that the curves match pretty well, so continuity of the derivative (where there is no mirror) is satisfied. This matching condition can be improved by taking more values of $\ell$. However, we see that the used cutoff already gives a good approximation of the analysed system.

\subsection{Resonance on the electromagnetic cavity}\label{appendix:resonance}

\begin{figure}[t]
 \centering
    \includegraphics[width=.8\linewidth]{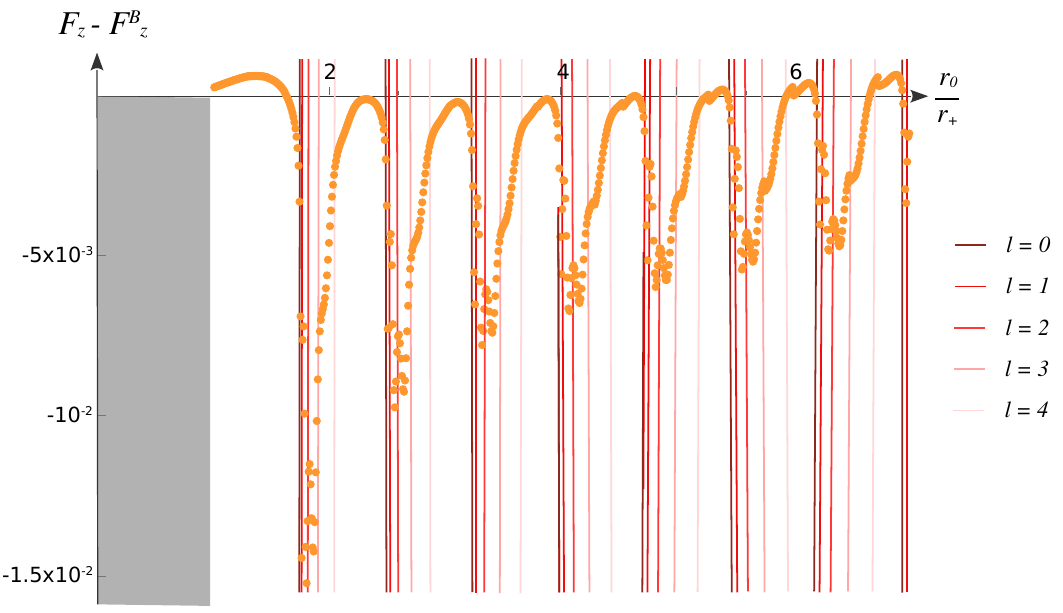}
    \caption{Total thrust force at $k=0.37$ obtained as function of the mirror radius $r_{\sf m}$ normalized with conductor sphere radius $r_+=11.4107$. Here we fix  ${\sf e}{\sf m}=5$, $T=0.07$, $\mu=-0.01$ and the radius $r_+$ correspond to a black hole horizon with $Q/M=0.99$. The colored lines correspond to the zeros of $u_\ell(r)$ defined in \eqref{eq:u_EM} and varying $\ell$.}
    \label{fig:background_resonance}
\end{figure}

In this section we study the case of a pure electromagnetic case where we change the black hole for a perfect conductor sphere of the same radius $r_+$, maintaining the spherical mirror. To obtain the solution in this case, we consider a combination of the spherical Bessel solutions (the exact radial solutions of \eqref{eq:EOMradial} in the case without the black hole) such that it vanishes at the conductor sphere:
\begin{equation}\label{eq:u_EM}
    u_\ell(r)=y_\ell(k r_+)j_\ell(k r)-j_\ell(k r_+)y_\ell(k r)
\end{equation}
In fig \ref{fig:background_resonance} we show the total thrust obtained for a fix value of $k$ (this is, without performing the integral in $k$) in the same way than the background force in Sec. \ref{sec:numerical_force} and using the previous radial function. Moreover, we show the zeros of the function \eqref{eq:u_EM} as a function of $r$ for different $\ell$. This zeros correspond for a resonant modes in the case of a closed cavity, then we see in \ref{fig:background_resonance} that the peaks coincide perfectly with the resonant modes where the lowest $\ell$ are dominant. As $\ell$ grows, they group together next at the right of the peaks. Therefore, we conclude that the peaks observed in Sec. \ref{sec:numerical_force} are a characteristic of the geometry and not due to the black hole, although these are modified by its presence.

\end{document}